\documentclass{cpbtex}
\usepackage{graphicx,amsmath,amssymb,epsfig,comment}
\def\PT{\cal PT}
\newcommand{\prt}{\partial}
\newcommand{\cG}{\mathcal{G}}
\newcommand{\cR}{\mathcal{R}}

\begin{document}
\title{{\bf 
 Optical nonlinearity in a strongly interacting Rydberg atom ensemble
}}
\author{Lu Qin$^{1,*}$, 
 Zeyun Shi$^{2,} \thanks{These authors contributed equally to this work.}$,  
Xingdong Zhao$^{1}$, 
 Chao Hang$^{3,6,7,}$\thanks{Corresponding author. E-mail:~chang@phy.ecnu.edu.cn}, 
 Weibin Li$^{4}$, 
 and Guoxiang Huang$^{5,6,7}$\thanks{Corresponding author. E-mail:~gxhuang@fyust.edu.cn} 
\\
{\small
$^{1}${School of Physics, Henan Normal University, Xinxiang 453007, China}
}\\  
{\small
$^{2}${School of Artificial Intelligence, Hubei University of Automotive Technology, Shiyan 442002, China}
}\\ 
{\small
$^{3}${Institute of Quantum Science and Precision Measurement, East China Normal University,}
}\\
{\small
{ Shanghai 200241, China}
}\\
{\small
$^{4}${School of Physics and Astronomy, University of Nottingham, Nottingham, NG7 2RD, United Kingdom}
}\\
{\small
$^{5}${School of Arts and Sciences, Fuyao University of Science and Technology, Fuzhou 350109, China}
}\\
{\small
$^{6}${NYU-ECNU Institute of Physics, New York University Shanghai, Shanghai 200062, China} 
}\\
{\small
$^{7}${Collaborative Innovation Center of Extreme Optics, Shanxi University, Taiyuan 030006, China}
}
}

\date{\small{\today}}

\maketitle
\begin{abstract}

Rydberg atoms, characterized by their giant polarizability and strong long-range interactions, provide an ideal platform for exploring optical nonlinearities. By coupling light to ensembles of Rydberg atoms, one can map the strong and nonlocal interactions between Rydberg atoms and light fields. A particularly important scheme for realizing such nonlinearity is electromagnetically induced transparency (EIT).
This article presents a review of the research progress in nonlinear optics based on Rydberg-EIT systems, particularly nonlinear light propagation in Rydberg atomic gases. We first introduce the fundamental properties of Rydberg atoms and the physical mechanism of EIT-enhanced nonlinearity, emphasizing how the Rydberg blockade effect endows the system with a giant nonlocal Kerr nonlinearity. Subsequently, we discuss some typical nonlinear optical phenomena in such a system with focusing and defocusing nonlocal nonlinearities. In the focusing regime, we discuss spatial solitons, vortex solitons, soliton cloning, soliton molecules, optical bullets, and self-induced transparency solitons. Other topics are also discussed. 
In the defocusing regime, we analyze pattern formation and shock wave dynamics. Finally, we provide an outlook on the future development of this field toward quantum control, hybrid-system integration, and novel functional devices.
\end{abstract}



\tableofcontents

\section{Introduction}

Rydberg atoms are atoms in highly excited electronic states with a large principal quantum number $n$~\cite{Gallagher2008,saffman_quantum_2010,stebbings1983rydberg}. The weak Coulomb interaction between the nucleus and the Rydberg electron renders such Rydberg states very sensitive to their environment and, for instance, gives rise to an electric polarizability that scales as $n^7$~\cite{wu_concise_2021,Adams2020}. This leads to strong interactions between Rydberg atoms, which are key in many quantum simulations, computations, and sensing applications, which are very active research areas~\cite{Browaeys2020,Saffman2010,Adams2020}. 
This large polarizability has been exploited to realize a drastically enhanced Kerr effect in cold Rydberg gases ~\cite{mohapatra_giant_2008}. 
The giant Kerr nonlinearity arising from Rydberg interactions provides a powerful tool for manipulating light at the few-photon level~\cite{Baur2014,Gorniaczyk2014,Chang2014}, enabling a variety of nonlinear optical phenomena that are inaccessible in conventional media~\cite{Firstenberg2016}.

A key phenomenon underlying this giant nonlinearity is the Rydberg blockade effect. When two atoms are excited to Rydberg states within a critical distance known as the blockade radius, the interaction-induced energy shift exceeds the spectral width of the excitation laser, preventing the simultaneous excitation of both atoms~\cite{Urban2009NP}. This effect enables the control of collective excitations in an atomic ensemble, where the entire ensemble behaves as a single superatom that can host at most one Rydberg excitation. The Rydberg blockade has become a cornerstone of many quantum technologies. In quantum computation, it provides a mechanism for implementing high-fidelity two-qubit gates between neutral atoms~\cite{Jaksch2000,Saffman2010}. Quantum simulation allows the realization of spin models, such as the Ising and XY models, by mapping Rydberg excitations onto effective spin degrees of freedom~\cite{Schauss2018}. Most relevant to this review, the blockade effect can be mapped onto optical fields via electromagnetically induced transparency (EIT), yielding a giant nonlocal Kerr nonlinearity. Specifically, a single Rydberg excitation can block the excitation of subsequent photons, leading to phenomena such as photon blockade and single-photon switches~\cite{Baur2014,Gorniaczyk2014}. These capabilities position Rydberg-blockaded ensembles as a powerful platform for exploring strongly correlated photonic physics and advancing quantum information-processing.

While several existing review papers have comprehensively surveyed the broad landscape of Rydberg atom technologies~\cite{Browaeys2020,Saffman2016,Schauss2018,Adams2020,Beterov2020,Henriet2020,Morgado2021,Wu2024,Shi2022,Yuan2023}—covering their applications in quantum computation, simulation, and sensing—the specific area of Rydberg-mediated optical nonlinearity, particularly how it affects the propagation dynamics and stability of light fields, remains relatively underexplored. This review article distinguishes itself by focusing specifically on the realization of optical nonlinearity in a strongly interacting Rydberg atom ensemble and its impact on light field propagation and stability.


\subsection{Nonlinear optics}

Photons, as mediators of electromagnetic interactions, are devoid of direct mutual interactions in a vacuum. Although this property renders light an ideal carrier for information transmission, it simultaneously imposes fundamental limitations on the exploitation of optical fields in quantum information processing and nonlinear optical control. The realization of controllable photon-photon interactions, regardless of whether within the classical or quantum regime, constitutes a core objective of contemporary optical research~\cite{Murray2016}.

Traditional nonlinear optics rely on the nonlinear response of a medium to an optical field. However, the Kerr nonlinearity of conventional materials is generally weak, requiring intense laser fields to generate nonlinear effects that are observable. To realize nonlinear effects at the few-photon or even single-photon level, researchers have pursued two primary strategies: the first involves coupling photons to single quantum emitters (such as single atoms or quantum dots) and enhancing the local optical field strength using high-Q optical microcavities~\cite{vahala2003optical,reithmaier2004strong,haroche2006exploring}. The second strategy leverages the collective enhancement of an atomic ensemble combined with strong interatomic interactions to induce effective interactions between photons~\cite{Murray2016}. The core physical platform for the latter strategy is the Rydberg atomic ensemble.

\subsection{Rydberg atoms}

Rydberg atoms are atoms excited to high-lying electronic states (principal quantum number $n \gg 1$). Because the Bohr radius scales as $\sim n^2$, Rydberg atoms possess macroscopic atomic dimensions (up to micrometer scale) and a suite of dramatically enhanced physical properties: polarizability increases sharply as $\sim n^7$, the van der Waals interaction coefficient scales as $\sim n^{11}$, and the spontaneous emission lifetime is extended by $\sim n^3$, see Table~\ref{TAB0}. These attributes enable strong interactions between Rydberg atoms over macroscopic distances of tens of micrometers.

\begin{table}[h]
\renewcommand\tabcolsep{18pt}
\centering
\caption{\small Scaling with the principle quantum number $n$ for the listed properties of Rydberg atoms.
}
\vspace{0.1cm}
\label{TAB0}
\begin{tabular}{ccc}
\hline\hline\vspace{-0.5cm}&\\
Properties &
Notation & Scaling\\
\hline
&\vspace{-0.5cm}\\
orbit size
& $\langle r\rangle$  & $n^{2}$   \\
polarizability 
& $\alpha$ & $n^{7}$ \\
van der Waals interaction coefficient 
& $C_6$ & $n^{11}$ \\
radiative lifetime 
& $\tau_0$ & $n^{3}$ \\
\hline\hline
\end{tabular}
\end{table}

\begin{figure}[ht]
\centering
\includegraphics[width=0.45\linewidth]{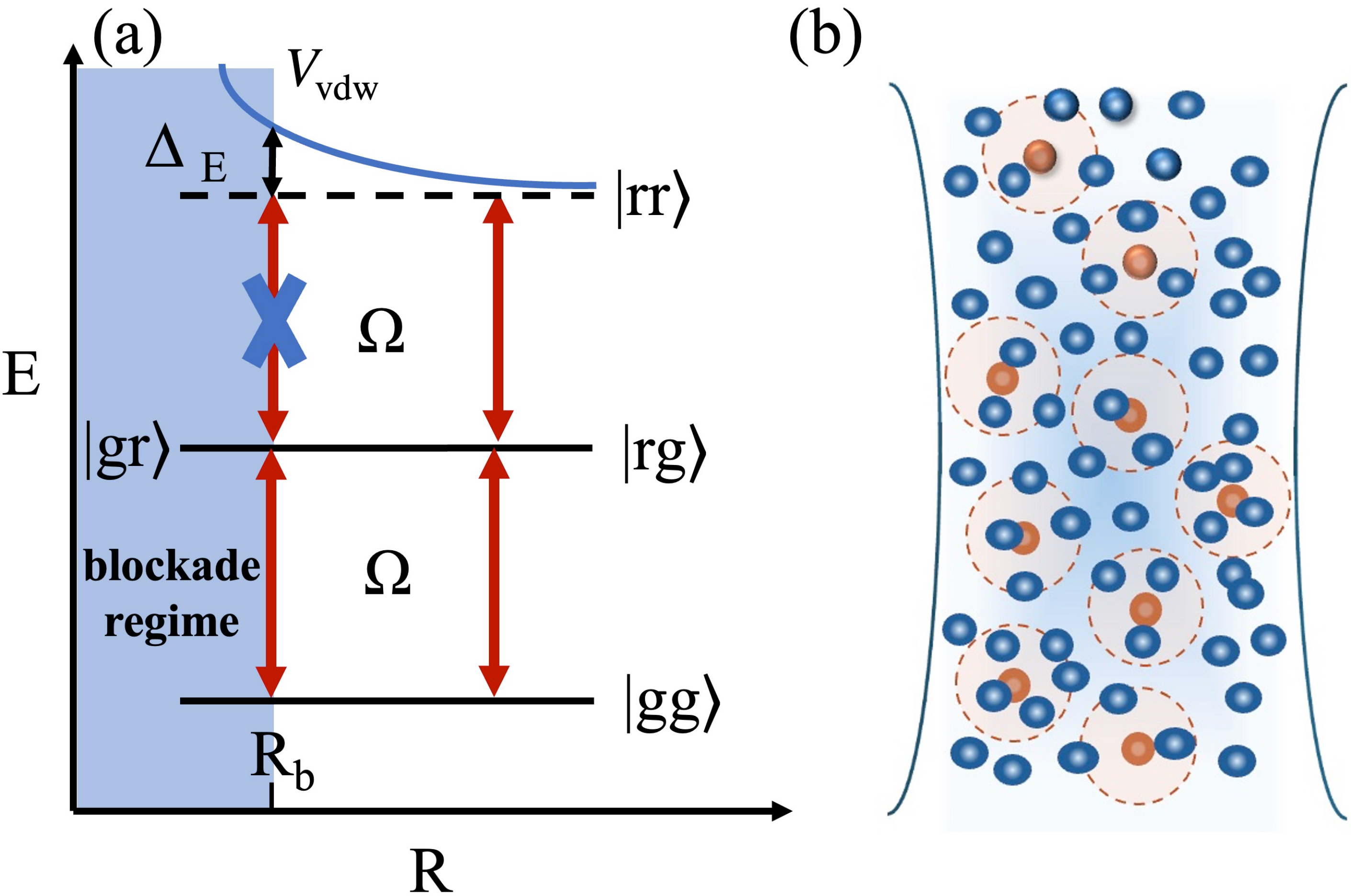}
\caption{\footnotesize 
Rydberg interaction and Rydberg blockade.  (a) Rydberg blockade in the two-atom case. (b) Rydberg superatoms in an ensemble of atoms. All atoms inside a volume (red dashed circle) with radius $R_b$ share a single Rydberg excitation (red sphere).
}
\label{ch1_blockade}
\end{figure}

An important property of Rydberg physics is the Rydberg blockade effect. When the excitation light attempts to excite an atom from the ground state to a Rydberg state, if one atom is already excited, its strong interaction significantly shifts the Rydberg energy levels of the neighboring atoms, as shown in Fig.~\ref{ch1_blockade}(a). Within a blockade radius $R_b$, the interaction-induced level shift exceeds the linewidth of the excitation laser, thereby suppressing the generation of a second Rydberg excitation. This implies that within the blockade volume, at most one Rydberg excitation can exist, forming a collective ``superatom" state; see Fig.~\ref{ch1_blockade}(b). This effect lays a physical foundation for mapping strong interatomic interactions onto photonic states. In addition, using two-body dephasing, an enhanced blockade effect can be achieved~\cite{Yan:20}.

\subsection{Nonlinear optics with Rydberg atoms}

EIT is a coherent optical phenomenon that utilizes quantum interference to eliminate medium absorption~\cite{Fleischhauer2005}. Combining EIT with Rydberg atoms, specifically by replacing the upper state in a conventional ladder-type EIT configuration with a highly excited Rydberg state, creates a unique platform for nonlinear optics~\cite{Mohapatra2007}. In this configuration, the probe light propagates with minimal loss and an extremely slow group velocity within the transparency window created by the control light, while the Rydberg blockade effect simultaneously imparts a giant equivalent optical Kerr nonlinearity to the system.

The nonlinearity of the Rydberg-EIT system has two key characteristics. The first is nonlocality: since the blockade radius $R_b$ can be much larger than the optical wavelength, the Rydberg excitation generated by a photon at one location affects the optical response of atoms at distant locations, endowing the nonlinear response function with long-range spatial correlations~\cite{sevinccli2011PRL}. The second is tunability: by selecting distinct atomic species and tuning the Rydberg state ($S,\,P,\,D,\cdots$), one can switch between focusing (self-focusing) and defocusing (self-defocusing) nonlocal nonlinearities and precisely tune the strength and range of nonlinearity. These two characteristics make the Rydberg-EIT system an ideal experimental and theoretical platform for investigating a variety of nonlinear wave phenomena, from solitons and vortices to pattern formation and shock waves.

\section{Physical model}
\subsection{The physical setup}

\begin{figure}[ht]
\centering
\includegraphics[width=0.6\linewidth]{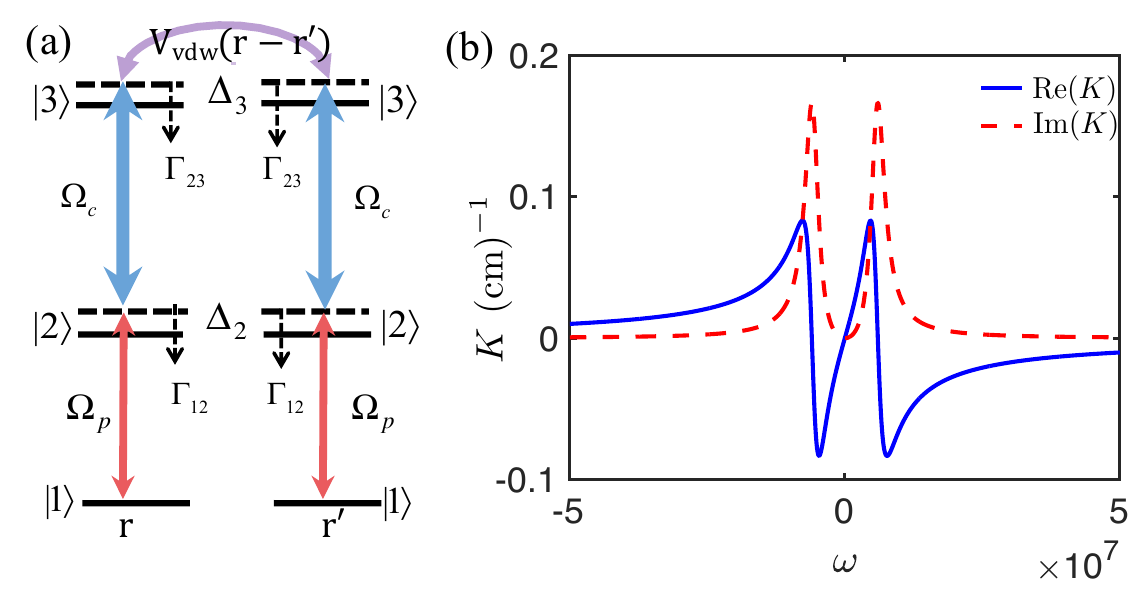}
\caption{\footnotesize 
(a)~Schematic of the laser-cooled, dilute three-level Rydberg-EIT atomic gas. A weak probe laser field $\mathbf{E}_{p}$ with half-Rabi frequency $\Omega_{p}$ drives the $|1\rangle \leftrightarrow |2\rangle$ transition, while a strong continuous-wave control field $\mathbf{E}_{c}$ with half-Rabi frequency $\Omega_{c}$ drives the $|2\rangle \leftrightarrow |3\rangle$ transition. The states $|1\rangle$, $|2\rangle$, and $|3\rangle$ denote the ground, intermediate, and high-lying Rydberg states, respectively. 
$\Gamma_{12}$ and $\Gamma_{23}$ are spontaneous emission decay rates. The interaction between two Rydberg atoms located at positions $\mathbf{r}$ and $\mathbf{r}'$ is described by the vdW potential.
(b) Linear dispersion relation of probe field. Real~(blue solid line) and imaginary~(red dashed line) parts of the linear dispersion relation as functions of the central frequency $\omega$.
}
\label{ch2_model}
\end{figure}

A typical theoretical model for Rydberg-EIT nonlinear optics, based on a three-level atomic system (e.g. rubidium, strontium), is shown in Fig.~\ref{ch2_model}(a)~\cite{Bai2016OE,Bai2019,qin2022stable}. Here a weak probe laser field $\mathbf{E}_{p}$ with half-Rabi frequency $\Omega_{p}$, couples to the transition $|1\rangle \leftrightarrow |2\rangle$, and a strong, continuous-wave control laser field $\mathbf{E}_{c}$ with half-Rabi frequency $\Omega _{c}$, couples to the transition $|2\rangle\leftrightarrow |3\rangle $, where $|1\rangle $, $|2\rangle $, and $|3\rangle $ denote, respectively, the ground, intermediate, and high-lying Rydberg states; $\Gamma_{12}$ and $\Gamma_{23}$ are the spontaneous emission decay rates. The interaction between the two Rydberg atoms respectively at positions $\mathbf{r}$ and $\mathbf{r}^{\prime }$ is described by the long-range van der Waals (vdW) potential~\cite{pritchard2010cooperative}
\begin{eqnarray}
V_{\mathrm{vdW}}=\hbar V(\mathbf{r}^{\prime }-\mathbf{r})\equiv -\frac{\hbar
C_{6}}{|\mathbf{r^{\prime }}-\mathbf{r}|^{6}},  \label{vdW}
\end{eqnarray}
with $C_6$ as the dispersion coefficient. 

The Hamiltonian of the system is given by $\hat{H}=\mathcal{N}_a\int d^3{\bf r}\hat{\mathcal{H}}_0({\bf r},t)+(\mathcal{N}_a/2)\int d^3{\bf r}\hat{\mathcal{H}}_1({\bf r},t)$. Here 
$\mathcal{N}_a$ is the atomic density, $\hat{\mathcal{H}}_0({\bf r},t)$ describes the atoms and the coupling between the atoms and light fields, $\hat{\mathcal{H}}_1({\bf r},t)$ describes the Rydberg-Rydberg interaction. Under the electric-dipole and rotating-wave approximations, $\hat{\mathcal{H}}_0$ and $\hat{\mathcal{H}}_1$ have the forms
\begin{eqnarray}
&& \hat{{\mathcal H}}_0= -\sum_{\alpha =2}^{3}{\hbar \Delta _{\alpha }\hat{S}%
_{\alpha \alpha }\left( \mathbf{r},t\right) }-\hbar \left[ \Omega _{p}\hat{S}%
_{12}+\Omega _{c}\hat{S}_{23}+\mathrm{h.c.}\right], \nonumber \\
&& \hat{\mathcal{H}}_1=\mathcal{N}_{a} \int{d^3 {\bf r}^{\prime}\hat{S}_{33}({\bf r}',t) \hbar V ({\bf r}^{\prime}-{\bf r}) \hat{S}_{33} ({\bf r},t )}.\notag \label{Hamiltonian1}
\end{eqnarray}
Here $\Delta_{2}$ and $\Delta_{3}$ are, respectively, the one- and
two-photon detunings; 
$\Omega _{p}={(\mathbf{e}_{p}\cdot \mathbf{p}_{12})\mathcal{E}_{p}}/(2\hbar )$
and $\Omega_{c}={(\mathbf{e}_{c}\cdot \mathbf{p}_{23})\mathcal{E}_{c}}%
/(2\hbar )$ are, respectively, half Rabi frequencies of the probe and
control fields, with $\mathbf{p}_{\alpha \beta }$ the electric-dipole matrix
elements associated with the transition $|\beta\rangle \leftrightarrow
|\alpha \rangle $; $\hat{S}_{\alpha \beta }\equiv|\beta\rangle\langle\alpha|\exp\{i[(\mathbf{k}_{\beta }-\mathbf{k}_{\alpha })\cdot
\mathbf{r}-(\omega_{\beta }-\omega_{\alpha }+\Delta_{\beta }-\Delta
_{\alpha})t]\}$ is the atomic transition operator $(\alpha,\,\beta=1,\,2,\,3)$, satisfying the commutation relation
$[\hat{S}_{\alpha\beta}({\bf
r},t),\hat{S}_{\alpha^\prime\beta^\prime}({\bf r}^\prime,t)]
=\mathcal{N}_a^{-1}
\delta ({\bf r}-{\bf r}{^\prime})[\delta_{\alpha\beta'}\hat{S}_{\alpha'\beta}({\bf r},t)-\delta_{\alpha'\beta}\hat{S}_{\alpha\beta'}({\bf r'},t)]$,
with $\delta_{\alpha\beta}$ being the Kronecker symbol.

The atomic dynamics is governed by the Heisenberg equation of motion for the operators
$\hat{S}_{\alpha\beta}({\bf r},t)$, i.e. $i\hbar{\partial}\hat{S}_{\alpha\beta}({\bf r},t)/{\partial t}=[\hat{H}, \hat{S}_{\alpha\beta}({\bf r},t)]$. Taking expectation values on the both sides of this equation, we obtain the optical Bloch equation involving one- and two-body reduced density matrices, with the form~\cite{qin2022stable}
\begin{eqnarray}\label{Bloch0}
\frac{\partial\hat{\rho}}{\partial t}=-\frac{i}{\hbar} \left[{\hat{ H}_{0}},\hat{\rho}\right]-\Gamma\left[\hat{\rho}\right]+\hat{R}\,[\hat{\rho}_{\rm 2body}],
\end{eqnarray}
where $\hat{\rho} ({\bf r},t)$ is reduced one-body  density matrix (DM) with matrix elements  $\rho_{\alpha\beta}({\bf r},t)\equiv\langle \hat{S}_{\alpha\beta} ({\bf r},t)\rangle$, $\Gamma$ is a $3\times 3$
relaxation matrix describing spontaneous emission and dephasing. Due to the existence of the Rydberg-Rydberg interaction, two-body reduced DM is involved in Eq.~(\ref{Bloch0}), represented by the last term $\hat{R}\,[\hat{\rho}_{\rm 2body}]$. The full system of optical Bloch equations for the evolution of the
density-matrix elements $\rho _{\alpha \beta }$ is
\begin{subequations}
\label{Bloch}
\begin{align}
& i\frac{\partial }{\partial t}\rho _{11}-i\Gamma _{12}\rho _{22}+\Omega
_{p}^{\ast }\rho _{21}-\Omega _{p}\rho _{12}=0,  \label{Bloch11} \\
& i\frac{\partial }{\partial t}\rho _{22}+i\Gamma _{12}\rho _{22}-i\Gamma
_{23}\rho _{33}+\Omega _{c}^{\ast }\rho _{32}-\Omega _{c}\rho _{23}-\Omega
_{p}^{\ast }\rho _{21}+\Omega _{p}\rho _{12}=0,  \label{Bloch22} \\
& i\frac{\partial }{\partial t}\rho _{33}+i\Gamma _{23}\rho _{33}-\Omega
_{c}^{\ast }\rho _{32}+\Omega _{c}\rho _{23}=0,  \label{Bloch33} \\
& \left( i\frac{\partial }{\partial t}+d_{21}\right) \rho _{21}+\Omega
_{c}^{\ast }\rho _{31}-\Omega _{p}(\rho _{22}-\rho _{11})=0,  \label{Bloch21}
\\
& \left( i\frac{\partial }{\partial t}+d_{31}\right) \rho _{31}-\Omega
_{p}\rho _{32}+\Omega _{c}\rho _{21}-\frac{\mathcal{N}%
_{a}}{2}\int d^{3}\mathbf{%
r^{\prime }}V(\mathbf{r^{\prime }}-\mathbf{r})\rho \rho _{33,31}(\mathbf{%
r^{\prime }},\mathbf{r},t)=0,  \label{Bloch31} \\
& \left( i\frac{\partial }{\partial t}+d_{32}\right) \rho _{32}-\Omega
_{p}^{\ast }\rho _{31}-\Omega _{c}(\rho _{33}-\rho _{22})-\frac{\mathcal{N}%
_{a}}{2}\int d^{3}\mathbf{r^{\prime }}V(\mathbf{r^{\prime }}-\mathbf{r})\rho
\rho _{33,32}(\mathbf{r^{\prime }},\mathbf{r},t)=0,  \label{Bloch32}
\end{align}%
\end{subequations}
where 
$d_{\alpha\beta}=\Delta_{\beta}-\Delta_{\alpha}+i\gamma_{\alpha\beta}$~($\Delta_1=0$; $\alpha\ne \beta$), $\gamma_{\alpha \beta }=(\Gamma _{\alpha }+\Gamma _{\beta
})/2+\gamma_{\alpha \beta }^{\mathrm{dep}}$, and $%
\Gamma_{\beta}=\sum_{\alpha <\beta }\Gamma _{\alpha \beta }$, with $\Gamma
_{\alpha \beta}$ and and $\gamma_{\alpha \beta }^{\mathrm{dep}}$ being the
spontaneous decay and dephasing rates, respectively.
In the last term on left-hand side of Eqs.~(\ref{Bloch31}) and (\ref{Bloch32}), the
two-body correlators $\rho _{33,3\alpha }(\mathbf{r^{\prime },r},t)\equiv
\langle {\hat{S}}_{33}(\mathbf{r^{\prime }},t){\hat{S}}_{3\alpha }(\mathbf{r}%
,t)\rangle $ $(\alpha =1,\,2)$ originate from the Rydberg-Rydberg interaction.
The dynamical equations for the two-body correlators $\langle {\hat{S}}_{33}{\hat{S}}_{31}\rangle $ and $\langle {\hat{S}}_{33}{\hat{S}}_{32}\rangle $ are written as~\cite{Bai2016OE}:
\begin{subequations}
\label{correlators}
\begin{align}
& \left( i\frac{\partial }{\partial t}+d_{31}+i\Gamma _{23}-V(\mathbf{%
r^{\prime }-r})\right) \langle {\hat{S}}_{33}{\hat{S}}_{31}\rangle -\Omega
_{c}^{\ast }\langle {\hat{S}}_{32}{\hat{S}}_{31}\rangle +\Omega _{c}\left(
\langle {\hat{S}}_{23}{\hat{S}}_{31}\rangle +\langle {\hat{S}}_{33}{\hat{S}}%
_{21}\rangle \right)   \nonumber \\
& \qquad -\Omega _{p}\langle {\hat{S}}_{33}{\hat{S}}_{32}\rangle -\frac{\mathcal{N}%
_{a}}{2}\int d^{3}{{\rm r}^{\prime \prime }}\langle {\hat{S}}_{33}(\mathbf{r^{\prime
\prime }},t){\hat{S}}_{33}(\mathbf{r^{\prime }},t){\hat{S}}_{31}(\mathbf{r}%
,t)\rangle V(\mathbf{r^{\prime \prime }-r})=0, \\
& \left( i\frac{\partial }{\partial t}+d_{32}+i\Gamma _{23}-V(\mathbf{%
r^{\prime }-r})\right) \langle {\hat{S}}_{33}{\hat{S}}_{32}\rangle +\Omega
_{c}\left( \langle {\hat{S}}_{23}{\hat{S}}_{32}\rangle -\langle {\hat{S}}%
_{33}{\hat{S}}_{33}\rangle +\langle {\hat{S}}_{33}{\hat{S}}_{22}\rangle
\right)   \nonumber \\
& \qquad -\Omega _{p}^{\ast }\langle {\hat{S}}_{33}{\hat{S}}_{31}\rangle
-\Omega _{c}^{\ast }\langle {\hat{S}}_{32}{\hat{S}}_{32}\rangle -\frac{\mathcal{N}%
_{a}}{2}\int d^{3}{{\bf r}^{\prime \prime }}\langle {\hat{S}}_{33}(\mathbf{r^{\prime
\prime }},t){\hat{S}}_{33}(\mathbf{r^{\prime }},t){\hat{S}}_{32}(\mathbf{r}%
,t)\rangle V(\mathbf{r^{\prime \prime }-r})=0,
\end{align}%
\end{subequations}
where $\mathbf{r^{\prime }\neq r^{\prime \prime }}$ and $\hat{S}_{\alpha \beta }\hat{S}_{\mu \nu }$ in the terms without integration means $ \hat{S}_{\alpha \beta
}(\mathbf{r^{\prime }},t)\hat{S}_{\mu \nu }(\mathbf{r},t)$. 

These equations exhibit two features:
(i) The equations for two-body correlators  $\langle {\hat{S}}_{33}{\hat{S}}_{3\alpha}\rangle~(\alpha=1,\,2)$ involve many other two-body correlators (e.g. $\langle {\hat{S}}_{23}{\hat{S}}_{31}\rangle$, etc.). Thus, one must also solve additional equations for the other two-body correlators. 
(ii) The equations for the two-body correlators involve three-body correlators (e.g. $\langle {\hat{S}}_{33}{\hat{S}}_{33}{\hat{S}}_{31}\rangle$, etc.), which obey the equations of motion for three-body correlators (which are lengthy and not listed here) and also have to be solved. Similarly, the equations of motion of the three-body correlators involve the four-body correlators. Finally, an infinite hierarchy of equations of motion for the correlators of one, two, three, and so on, bodies is obtained.  To solve such equation chains, a suitable approximation is required. Using the second-order ladder approximation, such that for moderate atomic density the three-body correlation terms in the two-body correlator equations are factorized in the following way~\cite{sevinccli2011PRL,mukamel1995principles}
\begin{align}
&\langle {\hat{S}}_{\alpha \beta }(\mathbf{r}^{\prime \prime },t){\hat{S}}%
_{\mu \nu }(\mathbf{r^{\prime }},t){\hat{S}}_{\gamma \delta }(\mathbf{r}%
,t)\rangle  
=\langle {\hat{S}}_{\alpha \beta }(\mathbf{r^{\prime \prime }},t)\rangle
\langle {\hat{S}}_{\mu \nu }(\mathbf{r^{\prime }},t){\hat{S}}_{\gamma \delta
}(\mathbf{r},t)\rangle +\langle {\hat{S}}_{\alpha \beta }(\mathbf{r^{\prime
\prime }},t){\hat{S}}_{\mu \nu }(\mathbf{r^{\prime }},t)\rangle \langle {%
\hat{S}}_{\gamma \delta }(\mathbf{r},t)\rangle   \nonumber \\
& \qquad +\langle {\hat{S}}_{\alpha \beta }(\mathbf{r^{\prime \prime }},t){\hat{S}}%
_{\gamma \delta }(\mathbf{r},t)\rangle \langle {\hat{S}}_{\mu \nu }(\mathbf{%
r^{\prime }},t)\rangle -2\langle {\hat{S}}_{\alpha \beta }(\mathbf{r^{\prime
\prime }},t)\rangle \langle {\hat{S}}_{\mu \nu }(\mathbf{r^{\prime }}%
,t)\rangle \langle {\hat{S}}_{\gamma \delta }(\mathbf{r},t)\rangle.
\end{align} 



To simplify the hierarchy of equations, additional approximations can be adopted. In particular, when the atomic density is relatively low and the interaction between atoms is relatively weak, such that the correlation between atoms is negligible, one has $\langle S_{\alpha\beta}(\mathbf{r}'')S_{\mu\nu}(\mathbf{r}')S_{\alpha'\beta'}(\mathbf{r})\rangle \to \langle S_{\alpha\beta}(\mathbf{r}'')\rangle\langle S_{\mu\nu}(\mathbf{r}')\rangle\langle S_{\alpha'\beta'}(\mathbf{r})\rangle$, corresponding to a mean-field approximation (MFA). Another approach is the ground-state approximation (GSA), which assumes that the population in the atomic quantum states does not change during the time evolution of the system, for example, $\rho_{11}=1$ and $\rho_{22}=\rho_{33}=0$. With the GSA, the number of equations for the one-body DM elements is significantly reduced, and the calculation of the nonlinear optical response of the system can be greatly simplified. 
However, MFA and GSA are usually not valid for Rydberg gases, even at low atomic densities, owing to the strong Rydberg-Rydberg interaction. However, to acquire a giant nonlinear optical effect, a moderate atomic density is usually required; hence, one must adopt a method beyond the MFA and GSA, that is, the reduced density matrix expansion~\cite{Bai2016OE,Mu2021}.

\subsection{Solutions based on perturbation expansion}

To analytically obtain the nonlinear optical response of the system, we employ the reduced density matrix expansion developed in Refs.~\cite{Bai2016OE,Mu2021}. To this end, we assume that the atoms are initially populated in the state $|1\rangle$ and that the probe field is much weaker than the control field. Under such assumptions, we can make the asymptotic expansions
\begin{subequations}\label{expan}
\begin{eqnarray}
&& \rho_{\alpha\beta}=\rho_{\alpha\beta}^{(0)}+\epsilon
\rho_{\alpha\beta}^{(1)}+\epsilon^2 \rho_{\alpha\beta}^{(2)}+\epsilon^3 \rho_{\alpha\beta}^{(3)}\cdots,\,\,\,\, (\alpha, \beta=1,2,3; \,
\rho_{\alpha\beta}^{(0)}=\delta_{\alpha 1}\delta_{\beta 1}), \\
&& \rho_{\alpha\beta,\mu\nu}=\epsilon^2 \rho_{\alpha\beta,\mu\nu}^{(2)}+\epsilon^3 \rho_{\alpha\beta,\mu\nu}^{(3)}\cdots, \\
&& \Omega_p=\epsilon\Omega_{p}^{(1)}+\epsilon^2\Omega_{p}^{(2)}+\epsilon^3
\Omega_{p}^{(3)}+\cdots.
\end{eqnarray}
\end{subequations}
Here, \(\epsilon\) is a small dimensionless  parameter introduced to characterize the relative weakness of the probe field compared with the control field, \(|\Omega_p| \ll |\Omega_c|\). Consequently, the expansion in powers of \(\epsilon\) is equivalent to an expansion in powers of the probe-field amplitude.
Substituting the expansion into the
Bloch equations, we obtain a set of linear inhomogeneous equations that can be solved order by order.

Setting $\Omega_p= F\exp[i(Kz-\omega t)]$, with $F$ the slowly-varying envelope function, we obtain the solution at the first-order ($\sim \epsilon$) reading as
\begin{subequations}
\begin{align}
& \rho _{21}^{(1)}=\frac{(\omega +d_{31})}{D}\Omega _{p}\equiv
a_{21}^{(1)}\Omega _{p}, \\
& \rho _{31}^{(1)}=-\frac{\Omega _{c}}{D}\Omega _{p}\equiv
a_{31}^{(1)}\Omega_{p},
\end{align}%
\end{subequations}
where $D=|\Omega _{c}|^{2}-(\omega +d_{21})(\omega +d_{31})$, and other
matrix elements $\rho _{\alpha \beta }^{(1)}$ vanish. The linear dispersion relation for the probe field is given by
\begin{eqnarray}\label{eqn4}
K(\omega)=\frac{\omega }{c}+\kappa _{12}\frac{\omega+d_{31}}{D}.
\end{eqnarray}
Figure~\ref{ch2_model}(b) illustrates the real part of $K(\omega)$, Re$(K)$ (blue solid line) and the imaginary part of $K(\omega)$, Im$(K)$ (red dashed line), as functions of $\omega$. Clearly, a transparency window is opened in Im$(K)$ when $\Omega_c\gg\Omega_p$, and hence the probe field can propagate in the resonant atomic system with negligible optical absorption, which is a key characteristic of EIT.

At the second-order ($\sim \epsilon^2$) approximation, we obtain the solution $\rho _{\alpha 1}^{(2)}=a_{\alpha 1}^{(2)}\partial F/\partial t_{1}\exp (i\theta )$, $\rho_{32}^{(2)}=a_{32}^{(2)}|F|^{2}$, $\rho _{\beta \beta }^{(2)}=a_{\beta \beta }^{(2)}|F|^{2}$, where $(\alpha =2,\,3;~\beta =1,\,2,\,3)$ and $\theta =Kz-\omega t$. The expressions of the coefficients are
\begin{subequations}
\begin{align}
& a_{21}^{(2)}=\frac{i}{\kappa _{12}}\left( \frac{1}{V_{g}}-\frac{1}{c}\right) , \\
& a_{31}^{(2)}=-\frac{i(\omega+d_{31})}{\Omega _{c}^{\ast}D}-\frac{(\omega+d_{21})}{\Omega _{c}^{\ast }}a_{21}^{(2)}, \\
& a_{11}^{(2)}=\frac{[i\Gamma _{23}-2|\Omega _{c}|^{2}A_{1}]A_{2}-i\Gamma
_{12}|\Omega _{c}|^{2}A_{3}}{-\Gamma _{12}\Gamma _{23}-i\Gamma _{12}|\Omega
_{c}|^{2}A_{1}}, \\
& a_{33}^{(2)}=\frac{1}{i\Gamma _{12}}\left( A_{2}-i\Gamma
_{12}a_{11}^{(2)}\right) , \\
& a_{32}^{(2)}=\frac{1}{d_{32}}\left( -\frac{\Omega _{c}}{D}+2\Omega
_{c}a_{33}^{(2)}+\Omega _{c}a_{11}^{(2)}\right) ,
\end{align}%
\end{subequations}
where $V_{g}=(\partial K/\partial \omega )^{-1}$ is the group velocity of the envelope $F$ and the coefficients $A_{1}=1/d_{32}-1/d_{32}^{\ast}$, $A_{2}=(\omega+d_{31}^{\ast})/D^{\ast }-(\omega+d_{31})/D$, and $A_{3}=1/(D^{\ast
}d_{32}^{\ast })-1/(Dd_{32})$. 

With the above solutions, we can go to the third order ($\sim \epsilon^3$), where the solution of $\rho_{21}^{(3)}$ has the form as
\begin{eqnarray}\label{rho3}
\rho_{21}^{(3)}=r_{21}^{(3,\rm{loc})}|\Omega_{{p}}|^{2}\Omega_{{p}}
+\int d^3{\bf r}' r_{21}^{(3,\rm{nloc})}|\Omega_{{p}}({\bf r}^{\prime})|^{2}\Omega_{{p}}({\bf r}),
\end{eqnarray}
where the coefficients
\begin{subequations}
\begin{eqnarray}
&&a_{21}^{(3,\rm{loc})}=\frac{1}{D}\left[(\omega+d_{31})(2a_{11}^{(2)}+a_{33}^{(2)})+\Omega _{c}^{\ast }a_{32}^{(2)}\right],\notag\\
&&a_{21}^{(3,\rm{nloc})}=\frac{\mathcal{N}_{a}\Omega
_{c}^{\ast}}{D}a_{33,31}^{(3)}(\mathbf{r^{\prime }-r})V(\mathbf{r^{\prime }-r}),\notag
\end{eqnarray}
\end{subequations}
with
\begin{align}
a_{33,31}^{(3)}
\approx\frac{-2|\Omega_c|^2\Omega_c(2\omega+d_{21}+d_{31})/|D|^2}{2(\omega+d_{21})|\Omega_c|^2+M[2\omega+2d_{31}-V({\mathbf{r}'-\mathbf{r}})]},\nonumber
\end{align} 
and $M=|\Omega_c|^2-(\omega+d_{21})(2\omega+d_{21}+d_{31})$. Note that the solution $\rho_{21}^{(3)}$ given by Eq.~(\ref{rho3}) includes the locally and nonlocally nonlinear terms, contributed by the atom-light interaction and the interaction between two Rydberg atoms (i.e. the Rydberg-Rydberg interaction), respectively.


The optical Kerr effect is an important effect in the light-atom interaction system, especially the third-order nonlinear effect, which takes 
\begin{eqnarray}\label{eqn10}
\chi=\chi^{(1)}+\chi_{\rm{loc}}^{(3)}\left|\mathcal{E}_{p}\right|^{2}
+\chi_{\rm{nloc}}^{(3)}\left|\mathcal{E}_{p}\right|^{2},
\end{eqnarray}
where $\chi^{(1)}$, $\chi_{\rm loc}^{(3)}$, and $\chi_{\rm nloc}^{(3)}$ are respectively the first-order (linear), the third order local and nonlocal (nonlinear) optical susceptibilities of probe field, defined by
\begin{subequations}\label{eqn11}
\begin{align}
&\chi^{(1)}=\frac{\mathcal{N}_{a}\left|\mathbf{p}_{12}
\right|^{2}}{\varepsilon_{0} \hbar}\frac{d_{31}}{D},\notag\\
&\chi_{{\rm loc}}^{(3)}=\frac{\mathcal{N}_{a}\left|\mathbf{p}_{12}
\right|^{4}}{\varepsilon_{0} \hbar^{3}D}\left[d_{31}(2\rho_{11}^{(2)}+\rho_{33}^{(2)})+\Omega _{c}^{\ast }\rho_{32}^{(2)}\right],\notag\\
&\chi_{\rm{nloc}}^{(3)}=\frac{\mathcal{N}_{a}^2\left|\mathbf{p}_{12}
\right|^{4}}{\varepsilon_{0} \hbar^{3}}
\frac{\Omega
_{c}^{\ast}}{D}\int d^{3}\mathbf{r^{\prime }}\rho
_{33,31}^{(3)}(\mathbf{r^{\prime }-r})V(\mathbf{r^{\prime }-r}),\notag
\end{align}
\end{subequations}
where $\chi_{\rm loc}^{(3)}$ and $\chi_{\rm nloc}^{(3)}$ are contributed by the photon-atom interaction (without Rydberg atom) and the Rydberg-Rydberg interaction, respectively.
 A notable characteristic of the nonlocal Kerr nonlinear susceptibilities $\chi_{\rm nloc}^{(3)}$ is that they are proportional to $\mathcal{N}_{a}^2$, while the local Kerr nonlinear susceptibilities $\chi^{(3)}_{\rm loc}$ are proportional to $\mathcal{N}_{a}$. Due to the strong Rydberg-Rydberg interaction, the nonlocal nonlinear susceptibilities are generally three or four orders of magnitude larger than the local nonlinear susceptibilities.

To address a typical example, one candidate is the laser-cooled strontium $^{88}%
\mathrm{Sr}$ atoms, with atomic levels $|1\rangle =\left\vert
5s^{2}~^{1}S_{0}\right\rangle$, $|2\rangle =\left\vert
5s5p~^{1}P_{1}\right\rangle$, and $|3\rangle =\left\vert
5sns~^{1}S_{0}\right\rangle$. For the principal quantum number $n=60$, the
dispersion parameter $C_{6}\approx 2\pi \times 81.6\,\mathrm{GHz}\cdot
\mu \mathrm{m}^{6}$ (which implies the Rydberg-Rydberg interaction is attractive). The spontaneous emission decay rates are $\Gamma_{12}\approx 2\pi \times 32$\,MHz and $\Gamma_{23}\approx 2\pi\times 16.7$\,kHz, while the detunings are $\Delta _{2}=-2\pi\times 240$\,MHz and $\Delta_{3}=-2\pi\times 0.16$\,MHz. The density of the atomic gas is $\mathcal{N}_{a}=9\times 10^{10}~\mathrm{cm}^{-3}$, and the half Rabi frequency of the control field is
$\Omega_{c}=2\pi \times 20$\,MHz. Since $\Delta _{2}\gg \Gamma _{12},\,\Delta _{3}$, the system operates in a dispersive nonlinearity regime~\cite{Bai2019}.
We obtain the results of the third-order optical susceptibilities, given in Tab.~\ref{TAB1}.

\begin{table}
\renewcommand\tabcolsep{17pt}
\centering
\caption{\small  Third-order Kerr nonlinear optical susceptibilities $\chi_{{\rm loc}}^{(3)}$  and $\chi_{{\rm nloc}}^{(3)}$ in the dispersion regime.
}
\vspace{0.1cm}
\label{TAB1}
\begin{tabular}{cccc}
\hline\hline\vspace{-0.5cm}&\\
Susceptibility~$(\mathrm{m}^{2}\,\mathrm{V}^{-2})$ &
Real part & Imaginary part  & Contributed by\\
\hline
&\vspace{-0.5cm}\\
$\chi_{\rm{loc}}^{(3)}$
&$4.45\times10^{-10}$  & $5.63\times10^{-12}$  & photon-atom interaction \\
$\chi_{\rm{nloc}}^{(3)}$
& $2.32\times10^{-7}$ & $3.53\times10^{-9}$ & atom-atom interaction
\\
\hline\hline
\end{tabular}
\end{table}

From the table, we see that the Kerr nonlinear susceptibilities $\chi_{\rm{loc}}^{(3)}$ and  $\chi_{\rm{nloc}}^{(3)}$ possess the following interesting features:
(i)~Values of the real parts of the nonlocal nonlinear susceptibility can reach the order of magnitude  to $10^{-7}\,\mathrm{m}^2\,\mathrm{V}^{-2}$
for atomic density $\mathcal{N}_a=9\times10^{10}\,\mathrm{cm^{-3}}$. Such a giant Kerr nonlinearity originates from the strong Rydberg-Rydberg interaction in the system.
(ii)~The imaginary parts of the nonlinear susceptibilities (i.e., ${\rm{Im}}[\chi_{\rm{loc}}^{(3)}]$ and
${\rm{Im}}[\chi_{\rm{nloc}}^{(3)}]$) are much smaller than the corresponding real parts (i.e., ${\rm{Re}}[\chi_{\rm{loc}}^{(3)}]$ and
${\rm{Re}}[\chi_{\rm{nloc}}^{(3)}]$). Thus, nonlinear absorption in the system is largely suppressed, which is due to the EIT effect and the contribution of the large detuning $\Delta_2$.

For comparison with conventional nonlinear optical media, the third-order susceptibilities can also be expressed in terms of the nonlinear refractive index. Under the susceptibility convention adopted in this work, the conversion relation is~\cite{Boyd2008}
\[
n_2\,(\mathrm{m^2/W}) = \frac{3\chi^{(3)}}{4n_0^2 \epsilon_0 c},
\]
here $n_0$ is the linear refractive index, $\epsilon_0$ and $c$ are the vacuum permittivity and the speed of light in vacuum. For the dilute atomic gas considered here,  $n_0\simeq 1$. In Tab.~\ref{TAB_n} the exprimentally measured values of the nonlinear susceptibility are presented for several materials.

\begin{table}
\renewcommand\tabcolsep{17pt}
\centering
\caption{\small  The value of the nonlinear susceptibility and nonlinear refractive index, for various media.
}
\vspace{0.1cm}
\label{TAB_n}
\begin{tabular}{ccccc}
\hline\hline\vspace{-0.5cm}&\\
Medium & $n_0$ & $\chi^{(3)}$~($\mathrm{m^2/V^2})$ & $n_2~(\mathrm{m^2/W})$ & Reference\\
\hline
&\vspace{-0.5cm}\\
Vacuum & 1
&$3.4\times10^{-41}$  & $1\times10^{-38}$ & ~\cite{Euler1935}  \\
Air & 1.0003
&$1.7\times10^{-25}$  & $5\times10^{-23}$  & ~\cite{Pennington1989PRA}  \\
Water & 1.33
& $2.5\times10^{-22}$  & $4.1\times10^{-20}$ & ~\cite{Sutherland1996}\\
Glass (fused silica) & 1.47
& $2.5\times10^{-22}$  & $3.2\times10^{-20}$ & ~\cite{Chase1995}\\
Cold atom (BEC) & 1
& $7.1\times10^{-8}$  & $2\times10^{-5}$ &~\cite{Hau1999Nature}\\
Rydberg-EIT & 1
& $2.3\times10^{-7}$  & $6.5\times10^{-4}$ \\
\hline\hline
\end{tabular}
\end{table}

The third‑order truncation employed in theoretical framework is justified under the condition of low light intensity (low Rydberg excitation density), where higher‑order correlations are strongly suppressed due to the Rydberg blockade effect. In this regime, the infinite hierarchy of equations can be safely truncated at the third order.
However, Higher‑order corrections become necessary when the increased atomic density or higher probe intensities, which lead to a non‑negligible population of Rydberg states and thus significant fouth‑, or even higher order correlations. In such cases, the third‑order truncation is no longer sufficient, and one must include fifth‑order or higher terms in the perturbative expansion to capture the full quantum many‑body dynamics.

\subsection{Giant nonlocal self-Kerr and cross-Kerr nonlinearities}

Because cross-Kerr nonlinearities have potential applications ranging from optical quantum information processing to quantum nondemolition measurement, it is necessary to conduct a detailed theoretical and  experimental study of the Kerr effect in systems operating under double Rydberg-EIT.
Sinclair {\it et al.} reported the first experimental observation of the cross-Kerr nonlinearity in a cold $^{85}$Rb atomic gas with an inverted-Y-type level configuration via a double Rydberg EIT~\cite{Sinclair2019PRR}. By measuring the nonlinear phase written onto a probe laser pulse, they found that due to the Rydberg-Rydberg interaction the third-order nonlinear optical susceptibility $\chi^{(3)}$ of the system can be enhanced to the order of magnitude $1\times 10^{-8}\,{\rm cm}^{2}{\rm V}^{-2}$.

Using the reduced density matrix expansion~\cite{Mu2021}, we present systematic and detailed calculations of the third-order nonlinear optical susceptibilities of the system. 
We consider the dispersion regime by choosing
$\Delta_3=2\pi\times 100\,\mathrm{MHz}$, which for the cold $^{85}$Rb gives  $\Delta_3/\Gamma_3=16.5$. Other parameters are chosen to be
$\Delta_2=2\pi\times0.5\,\mathrm{MHz}$ (corresponding to $B=11.9\,\mu$T), $\Delta_4=2\pi\times 0.18\,\mathrm{MHz}$, $\Omega_{{c}}=2\pi\times 25\,\mathrm{MHz}$,
$\mathcal{N}_a=4\times10^{10}\,\mathrm{cm^{-3}}$, and $\rho_{11}^{(0)}=\rho_{22}^{(0)}=0.5$.
Because of the large detunings, the dephasing rates
$\gamma_{jl}^{\rm dep}$ can be neglected.
Based on the parameters, we obtain the result of the self-Kerr and cross-Kerr nonlinear optical susceptibilities, given in TABLE~\ref{TAB3}.

\begin{table}
\renewcommand\tabcolsep{17pt}
\centering
\caption{\small  Third-order Kerr nonlinear optical susceptibilities in the dispersion regime.  $\chi_{11}^{(3,{\rm nloc})}$  [$\chi_{22}^{(3,{\rm nloc})}$]: the third-order self-Kerr nonlinear susceptibility of the first (second) polarization component of the probe field. $\chi_{12}^{(3,{\rm nloc})}$ and $\chi_{21}^{(3,{\rm nloc})}$:  third-order cross-Kerr nonlinear susceptibilities between the two polarization components of the probe field.
{{System parameters used are
$\Delta_2=2\pi\times0.5\,\mathrm{MHz}$ with ${\rm B}=11.9\,{\rm{\mu T}}$, $\Delta_3=2\pi\times100\,\mathrm{MHz}$,
$\Delta_4=2\pi\times0.18\,\mathrm{MHz}$, $\Omega_{{c}}=2\pi\times25\,\mathrm{MHz}$}}, and $\mathcal{N}_a=4\times10^{10}\,\mathrm{cm^{-3}}$ (adapted from Ref.~\cite{Mu2021}).
}
\vspace{0.2cm}
\label{TAB3}
\begin{tabular}{ccc}
\hline\hline\vspace{-0.5cm}&\\
Susceptibility~$(\mathrm{m}^{2}\,\mathrm{V}^{-2})$ &
Real part & Imaginary part \\
\hline
&\vspace{-0.5cm}\\
$\chi_{11}^{(3,\rm{nloc})}$
&$-1.39\times10^{-8}$ &$1.82\times10^{-10}$\\
$\chi_{12}^{(3,\rm{nloc})}$
&$-1.21\times10^{-8}$ &$1.51\times10^{-10}$\\
$\chi_{21}^{(3,\rm{nloc})}$
&$-1.21\times10^{-8}$ &$2.12 \times10^{-10}$\\
$\chi_{22}^{(3,\rm{nloc})}$
&$-1.06\times10^{-8}$ & $1.79\times10^{-10}$
\\
\hline\hline
\end{tabular}
\end{table}

We show that such a system possesses giant nonlocal self-Kerr and cross-Kerr nonlinearities contributed by Rydberg-Rydberg interaction through double EIT. Our theoretical result on the cross-Kerr nonlinearity of the $^{85}$Rb atomic gas agrees with that of the experimental measurement by Sinclair {\it et al}.~\cite{Sinclair2019PRR}.
In addition, we demonstrate that the probe field may acquire a very large magneto-optical rotation (MOR) by applying  a very weak external magnetic field, which can be used to design atomic magnetometers with a precision much higher than that obtained using conventional EIT~\cite{Mu2021}. The research reported here opens a route for the development of nonlocal nonlinear and quantum optics based on Rydberg atoms and for practical applications in precision measurements, as well as in optical and quantum information processing and transmission~\cite{Pritchard2013in,Firstenberg2016,Murray2016}.

\section{Focusing nonlocal Kerr nonlinearity}

In traditional EIT systems, the nonlinearity is only local Kerr nonlinearity and remains very weak. 
 The combination of Rydberg atoms and EIT can overcome the drawback of weak nonlinearity in these systems. There are two types of Kerr nonlinearities in Rydberg EIT systems: the local optical Kerr nonlinearity arising from resonant light-atom interactions and the nonlocal optical Kerr nonlinearity arising from atom-atom interactions. When the atomic density increases, the latter can be significantly larger than the former.

\begin{figure}[ht]
\centering
\includegraphics[width=0.85\linewidth]{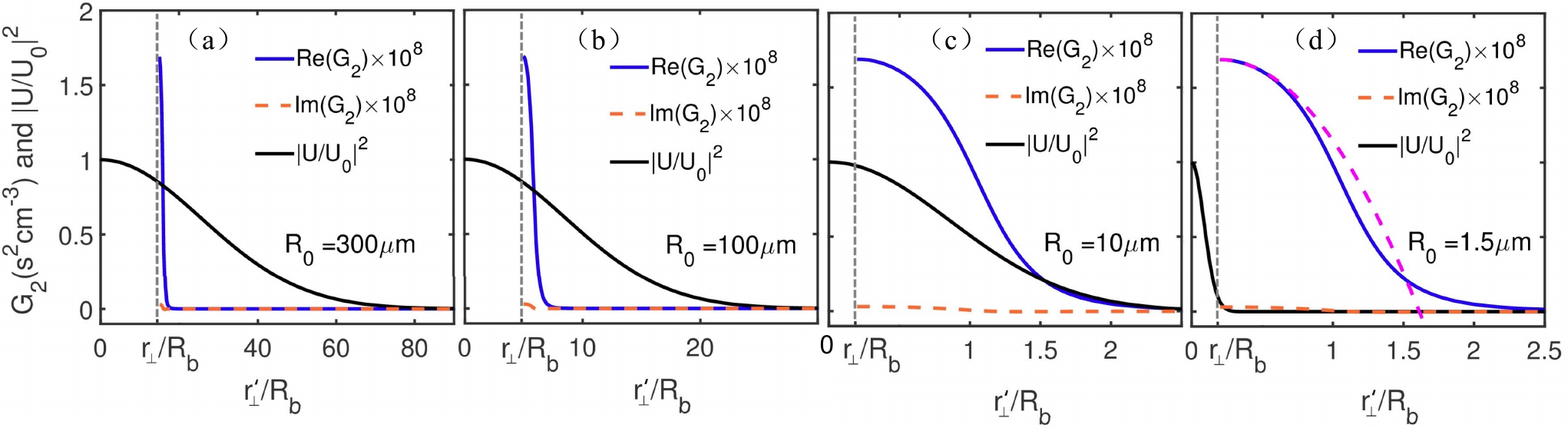}
\caption{\footnotesize  
Effective atomic interaction potential $G_2$ as a function of ${\rm r}_{\perp}'/R_b$.
(a)~Local response region with $R_0=300~\mu$m. (b)~Weak nonlocal response region with $R_0=100~\mu$m. (c)~Intermediate nonlocal response region with $R_0=10~\mu$m. (d)~Strong nonlocal response region with $R_0=1.5~\mu$m. Figures are adapted from Ref.~\cite{Bai2019}.
}
\label{ch1_Rydberg_nonlocal}
\end{figure}

The nonlocal Kerr nonlinear response in Rydberg atomic systems is determined by the ratio of the Rydberg blockade radius ($R_b$) to the beam waist radius ($R_0$) of the light field, i.e., $R_b/R_0$. 
When the range of the Rydberg-Rydberg interaction is much smaller than that of the beam radius of the probe pulse, i.e., $R_b/R_0\ll 1$, and the response function $G_2$ behaves similarly to a $\delta$-function. This means that for a large probe-field radius, the nonlocal Kerr effect contributed by the Rydberg-Rydberg interaction reduces into a local Kerr nonlinearity, see Fig.~\ref{ch1_Rydberg_nonlocal}(a).
When $|U|^2$ varies slowly compared with $G_2$, which corresponds to the weakly nonlocal response region, see Fig.~\ref{ch1_Rydberg_nonlocal}(b).
When $|U|^2$ exhibits a spatial variation similar to that of $G_2$, i.e. $R_b/R_0\sim 1$, the nonlinear optical response arising from the Rydberg-Rydberg interaction depends not only on the light intensity at a specific point, but also on the field distribution in its surrounding neighborhood, corresponding to the intermediately nonlocal response regime, see Fig.~\ref{ch1_Rydberg_nonlocal}(c).
When the range of the Rydberg–Rydberg interaction is much larger than that of the beam radius of the probe pulse, i.e. $R_b/R_0\gg 1$, which corresponds to strongly nonlocal response region, see Fig.~\ref{ch1_Rydberg_nonlocal}(d).

It has been shown that stable (2+1)D spatial optical (vortex) solitons and (3+1)D spatiotemporal optical solitons (also known as light bullets) can be created and stabilized in intermediately and strongly nonlocal regions. Moreover, through precise control of the interactions between solitons, stable bound states of (vortex) solitons, i.e, (vortex) soliton molecules, can also be formed.

\subsection{Spatial solitons and vortex solitons}

Solitons are self-trapped wave packets maintained by the interplay between dispersion (and/or diffraction) and nonlinearity of the host medium \cite{Zakharov,Ablowitz}. They are ubiquitous in nature, have been discovered in many areas, and have numerous applications. Owing to the largely suppressed optical absorption and strong nonlocal Kerr nonlinearity, a system with lasers and Rydberg atomic gas is a good platform for studying various high-dimensional (high-D) solitons and vortex solitons.

To derive the equation for the propagation of the probe field, we adopt the technique of multiple scales by introducing the multiscale variables $z_l=\epsilon^l z$ $(l=0,1,2)$, $x_1=\epsilon x$, $y_1=\epsilon y$, and $t_l=\epsilon^lt$ $(l=0,2)$. Consequently, $\rho_{\alpha\beta}^{(j)}$ and $\Omega_{p}^{(j)}$ ($j=1,2,3,...$) in the asymptotic expansions~(\ref{expan}) are considered functions of the multiscale variables. A solvability condition at the second order leads to the equation
\begin{align}
i\left(\frac{\partial F}{\partial z_1}+\frac{1}{V_g}\frac{\partial F}{\partial t_1}\right)=0,
\end{align}
which means the envelope $F$ propagates with the group velocity $V_g$.

At the third order, the solvability condition requires the envelope $F$ obeying another equation
\begin{align}\label{third_order}
& i\frac{\partial F}{\partial z_2}-\frac{K_2}{2}\frac{\partial^2 F}{\partial t_1^2}+\frac{1}{2 k_{p}}\left(\frac{\partial^2}{\partial x_1^2}+\frac{\partial^2}{\partial y_1^2}\right)F+W|F|^2Fe^{-2\bar{\alpha} z_2}
+Fe^{-2\bar{\alpha} z_2}\int d^3 {\bf r^{\prime}} G({\bf r^{\prime}-r}){|F({\bf r^{\prime}},t)|}^{2}=0.
\end{align}%
Here $\bar{\alpha}=\epsilon^{-2}{\rm Im}(K_0)$, the term with coefficient $W$ is contributed by the local Kerr nonlinearity, and the nonlocal response (integral kernel) function $G({\bf r^{\prime}-r})$ is contributed by the nonlocal Kerr nonlinearity due to the Rydberg-Rydberg interaction. The expressions of $W$ and $G$ are given by
\begin{subequations}\label{WG}
\begin{align}
& W=\kappa_{12}\frac{(\omega+d_{31})(a_{11}^{(2)}-a_{22}^{(2)})+\Omega_c^*a_{32}^{(2)}}{D}, \label{WG1}\\
& G({\bf r^{\prime}-r})=\kappa_{12}{\cal N}_a\frac{\Omega_c^*}{D}a_{33,31}^{(3)}({\bf r}'-{\bf r})V({\bf r'-r}), \label{WG2}
\end{align}%
\end{subequations}
respectively. Because $\kappa_{12}\propto {\cal N}_a$, one has $W \propto {\cal N}_a$ and $G({\bf r^{\prime}-r})\propto {\cal N}_a^2$. Thus, the magnitude of the nonlocal Kerr nonlinearity can be much larger than that of the local Kerr nonlinearity when the atomic density ${\cal N}_a$ has a typically moderate value (e.g. ${\cal N}_a\sim10^{10}$ cm$^{-3}$).

Equation~(\ref{third_order}) can be further written into a dimensionless form, reading as
\begin{align}\label{NNLSE0}
& i\left(\frac{\partial }{\partial \zeta }+a\right)u+d\frac{\partial
^{2}u}{\partial \tau ^{2}}+\left( \frac{\partial ^{2}}{\partial
\xi ^{2}}+\frac{\partial ^{2}}{\partial \eta ^{2}}\right) u+w|u|^{2}u +u\iint d\xi ^{\prime }d\eta ^{\prime }g(\xi ^{\prime }-\xi ,\eta ^{\prime
}-\eta )\left\vert u\left(\xi^{\prime},\eta^{\prime},\zeta,\tau\right) \right\vert
^{2}=0.
\end{align}%
Here, we have used the new variables $u=\Omega_p/U_0$, $\zeta=z/(2L_{\rm diff})$, $\tau=t/\tau_0$, $(\xi,\eta)=(x,y)/R_0$, with
$L_{\rm diff}=\omega_{p}R_0^2/c$ is the typical diffraction length, and $\tau_0$, $R_0$ and $U_0$ are the typical pulse duration, transverse size, and half Rabi frequency of the probe field, respectively. The coefficients $a=2L_{\rm diff}\bar{\alpha}$ and $d=L_{\rm diff}/L_{\rm disp}$, with $L_{\rm disp}=-\tau_0^2/K_2$ the typical dispersion length, the locally nonlinear coefficient $w=2L_{\rm diff}|U_0|^2W$, and the nonlocal response function $g(\xi ^{\prime }-\xi ,\eta ^{\prime
}-\eta )=2L_{\rm diff}R_0^2|U_0|^2 G({\bf r^{\prime}-r})$. Equation~(\ref{NNLSE0}) includes the linear absorption, group velocity dispersion, diffraction, local Kerr nonlinearity, and nonlocal Kerr nonlinearity, given by the first, second, third, fourth, and fifth terms on the left-hand side of the equation, respectively. For deriving Eq.~(\ref{NNLSE0}) we have assumed that the nonlocal character takes place in the transverse directions (i.e. $x$ and $y$ directions), which is valid only if the probe beam in the propagation direction (i.e. the $z$ direction) is rather extended. Equation~(\ref{NNLSE0}) has the form of a (3+1)D nonlocal nonlinear Schr\"{o}dinger equation (NNLSE).

Since the linear absorption can be largely suppressed via the EIT and the local Kerr nonlinearity is much less than the nonlocal one, further simplifications can be made so that Eq.~(\ref{NNLSE0}) is reduced as
\begin{align}\label{NNLSE1}
& i\frac{\partial u}{\partial \zeta }+\left(\frac{\partial ^{2}}{\partial
\xi ^{2}}+\frac{\partial ^{2}}{\partial \eta ^{2}}\right) u +u\iint d\xi ^{\prime }d\eta ^{\prime }g(\xi ^{\prime }-\xi ,\eta ^{\prime
}-\eta )\left\vert u\left(\xi^{\prime},\eta^{\prime},\zeta\right) \right\vert
^{2}=0.
\end{align}%
Here, we have also dropped the dispersion term ($d\propto 1/\tau_0^2 \approx0$, when $\tau_0^2\gg |K_2|L_{\rm diff}$), which is valid if the duration of the probe field is sufficiently long so that the dispersion effect becomes negligible.
 If the Kerr nonlinearity is a self-focusing one [i.e. $g(\xi ^{\prime }-\xi ,\eta ^{\prime}-\eta)>0$], the (2+1)D NNLSE~(\ref{NNLSE1}) can support stable 2D spatial solitons and vortex solitons due to the nonlocal character of the Kerr nonlinearity.

\begin{figure}[ht]
\centering
\includegraphics[width=0.45\linewidth]{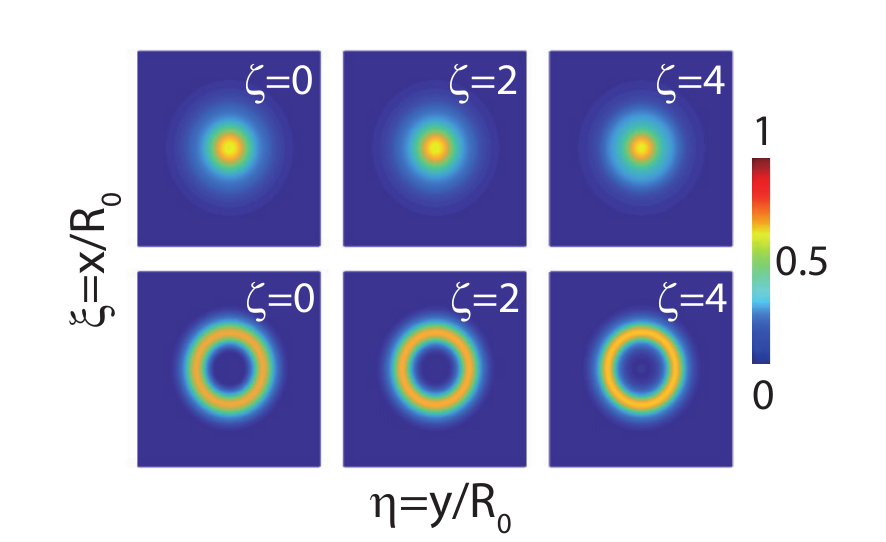}
\caption{\footnotesize The stable propagation of spatial solitons and vortex solitons. Upper row: Intensity of a typical spatial soliton as a function of $\xi=x/R_0$ and $\eta=y/R_0$ at $\zeta=z/(2L_{\rm diff})=0$, 2, and 4. Lower row: Same as for a typical vortex soliton. Figures are adapted from Ref.~\cite{qin2020PRA}.
}
\label{soliton_vortex}
\end{figure}

The upper row of Fig.~\ref{soliton_vortex} shows the intensity of a typical spatial soliton in the plane of $\xi$ and $\eta$ at $\zeta=0$ (i.e. the initial condition), $\zeta=2$ and $\zeta=4$, respectively, which does not exhibit obvious deformation during its propagation. The lower row of Fig.~\ref{soliton_vortex} shows the same result but for a typical spatial vortex soltion, which is also stable during its propagation.

\subsection{Soliton cloning}

\begin{figure}[t]
\centering
\includegraphics[width=0.8\linewidth]{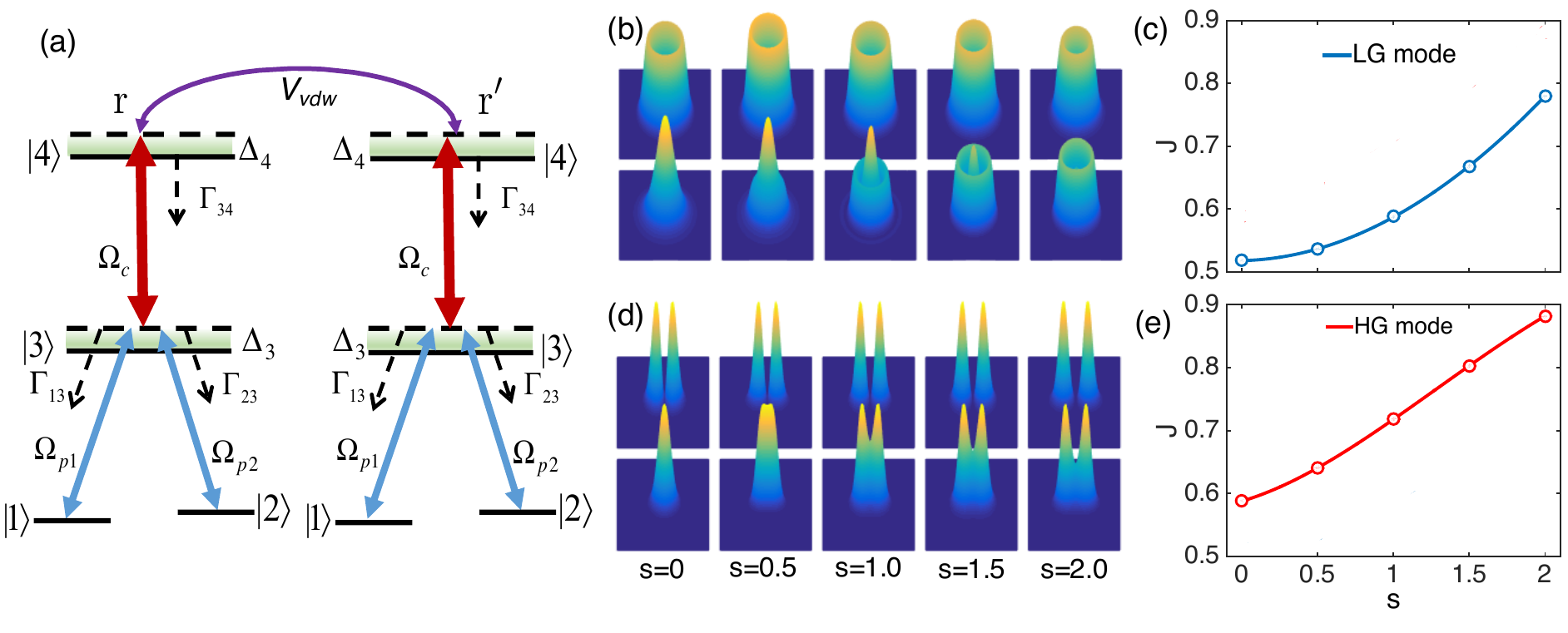}
\caption{\footnotesize {\bf Cloning of high-D optical beams}.
(a)~Inverted Y-type four-level configuration.
(b)~Cloning of optical vortex. Probe field 1 prepared as an optical vortex (upper row) is cloned onto probe beam 2 (hyperbolic secant mode; lower row). The normalized peak-intensity distributions of both probe fields as functions of the non-dimensional propagation distance are illustrated here.
(c)~Waveshape fidelity $J$ for cloning the Laguerre-Gaussian mode. 
(d) Same as (b) but for the cloning of Hermite-Gaussian mode (with $m=1$, $n=0$).
(e)Waveshape fidelity $J$ for cloning the Hermite-Gaussian mode. Figures are adapted from Ref.~\cite{qin2020PRA}.
}
\label{Ch2_Cloning}
\end{figure}

High-fidelity and controllable optical cloning of high-D optical beams is important for the development of novel techniques for optical imaging, lithography, and communications. In order to realize this, we consider a cold gas of Rydberg atoms with an inverted Y-type four-level configuration [see Fig.~\ref{Ch2_Cloning}(a)], where two weak probe laser fields with center frequencies $\omega_{p1}$ and $\omega_{p2}$ (half Rabi frequencies $\Omega_{p1}$ and $\Omega_{p2}$) couple the transitions $|1\rangle\leftrightarrow|3\rangle$ and $|2\rangle\leftrightarrow|3\rangle$, respectively. Here $|1\rangle$ and $|2\rangle$ are two ground states, and $|3\rangle$ is an intermediate state. In addition, a strong control laser field with center frequency $\omega_{c}$ (half Rabi frequency ${\Omega}_c$) couples the transition $|3\rangle\leftrightarrow|4\rangle$, where $|4\rangle$ is a high-lying Rydberg state. 

Note that in the inverted Y-shaped excitation scheme
There are two ladder-type excitation paths, that is, $|1\rangle \rightarrow|3\rangle \rightarrow|4\rangle$ and $|2\rangle \rightarrow|3\rangle \rightarrow|4\rangle$, which constitute two standard Rydberg-EITs (with the state $|4\rangle$ being a shared Rydberg state). Thus, the dynamics of the light fields and atoms in the system can be controlled by double Rydberg-EIT.

The propagation of the probe Rabi frequencies $\Omega_{p1}$ and $\Omega_{p2}$ obeys the following coupled nonlocal nonlinear Schr\"odinger equations (CNNLSE) in the dimensionless form 
\begin{subequations}\label{DNLS}
\begin{align}
&i\frac{\partial u}{\partial s}+\left(\frac{\partial ^{2}}{\partial
\xi ^{2}}+\frac{\partial ^{2}}{\partial \eta ^{2}}\right) u
+u\iint d\xi ^{\prime }d\eta ^{\prime }g_{11}(\xi ^{\prime }-\xi ,\eta ^{\prime
}-\eta )\left\vert u\left(\xi^{\prime},\eta^{\prime},s\right) \right\vert
^{2}=0,\label{DNLS1}\\
&i\frac{\partial v}{\partial s}+\left(\frac{\partial ^{2}}{\partial
\xi ^{2}}+\frac{\partial ^{2}}{\partial \eta ^{2}}\right) v
+v\iint d\xi ^{\prime }d\eta ^{\prime }g_{21}(\xi ^{\prime }-\xi ,\eta ^{\prime
}-\eta )\left\vert u\left(\xi^{\prime},\eta^{\prime},s\right) \right\vert
^{2}=0,\label{DNLS2}
\end{align}
\end{subequations}
where $(u,v)=(\Omega_{p1},\Omega_{p2})/U_0$, $s=z/(2L_{\rm diff})$, and $(\xi,\eta)=(x,y)/R_0$. Here
$L_{\rm diff}=\omega_{p}R_0^2/c$ is the typical diffraction length; $R_0$ and $U_0$ are the typical transverse size and half Rabi frequency of the probe fields, respectively. The dimensionless nonlocal response functions in the above equations are defined by $g_{\alpha 1}=2L_{\rm diff}R_0^2|U_0|^2G_{\alpha 1}$ ($\alpha=1,\,2$), where $G_{\alpha 1}$ has a similar form to Eq.~(\ref{WG2}). Assume that the intensity of probe field 1 is much stronger than that of probe field 2, and hence the cross-Kerr nonlinearity in Eq.~(\ref{DNLS1}) and the self-Kerr nonlinearity in Eq.~(\ref{DNLS2}) can be disregarded.

To illustrate the cloning of a (2+1)D optical vortex, the input intensities of probe field 1 and probe field 2 are respectively taken as the Laguerre-Gaussian mode (with the indexes $l=1$ and $p=0$) and Gaussian mode. The upper row of Fig.~\ref{Ch2_Cloning}(b) shows the normalized peak intensity at different propagation distances. Clearly, the wave shape of probe field 1 [controlled by Eq.~(\ref{DNLS1})] is cloned in probe field 2 [controlled by Eq.~(\ref{DNLS2})] when it propagates to a distance of $s=2.0$. From this figure, we find that probe field 2 eventually has a wave shape nearly the same as that of probe field 1.

The quality of the optical intensity cloning realized above can be characterized by the overlap integral of the input intensity of the probe field 1 (at position $s=0$) and the output intensity of the probe field 2 (at position $s$), which is called the wave shape fidelity $J$, defined by~\cite{Bai2019}
\begin{eqnarray}
&& J(s)=\frac{|\int_{-\infty}^{+\infty}d^2\zeta\,v(\vec{\zeta},s)\,
|u(\vec{\zeta},s=0)|\,|^2 }{\int_{-\infty}^{+\infty}d^2\zeta\,|v(\vec{\zeta},s)|^2  \int_{-\infty}^{+\infty}d^2\zeta\,|u(\vec{\zeta},s=0)|^2}.
\end{eqnarray}
Illustrated by the blue line in Fig.~\ref{Ch2_Cloning}(c)
is the wave shape fidelity $J$ as a function of $s$ for the cloning of the optical vortex, presented in Fig.~\ref{Ch2_Cloning}(b). At the propagation distance $s=2$, the wave shape of the optical vortex in probe field 1 can be well cloned in the wave shape of probe field 2 with high fidelity $J\approx 78\%$.

In Fig.~\ref{Ch2_Cloning}(d), 
The input intensities of probe fields 1 and 2 are taken as Hermite-Gaussian ($m=1$, $n=0$) and Gaussian modes, respectively.
The panel (e) of Fig.~\ref{Ch2_Cloning} gives the wave shape fidelity $J$ as a function of $s$ for the cloning, whose value is $88\%$ at $s=2.0$.

\subsection{Soliton molecules}

The bound states of solitons
\cite{Malomed1991,Malomed1993,Nepom,random,Afanasjev1997,Tang-1,Tang-2,Komarov2009,Wang2017,Skryabin,Gelash2019},
also known as soliton molecules (SMs), are objects of great interest as they demonstrate unique properties and offer various potential applications
in the area of photonics, including the design of new laser schemes, switchers, and data carriers~\cite{Yin2011,Rohrmann2013,Alamoudi2014,Kurtz2020}.

Based on Eq.~(\ref{NNLSE1}), a bound state of two solitons (i.e. a two-soliton molecule) can be sought by the ansatz $u(\xi ,\eta )=u_{+}(\xi ,\eta )+u_{-}(\xi ,\eta )$, with
\begin{equation}
u_{\pm }=A\,e^{-[\left(\xi \pm \mathcal{D}/{2}\right)^{2}+\eta^{2}]/(2a^{2})
+ib(\xi^{2}+\eta ^{2}) \pm i\pi/2}.\label{ansatz0}
\end{equation}
It includes two identical Gaussian beams, located at the positions $(-\mathcal{D}/2,0)$ and $(\mathcal{D}/2,0)$ (i.e. the separation of the two solitons is $\mathcal{D}$), with a $\pi$-phase difference. Here, $A$, $a$ and $b$ are the amplitude, width, and chirp of each soliton, respectively. The total power of the two-soliton molecule is $P=2\pi a^{2}A^{2}[1-e^{-\mathcal{D}^{2}/(4a^{2})}]$, which is conserved. Using the variational approach, we obtain a set of equations for $A$, $\mathcal{D}$, $b$, while $a$ can be determined by the total power with the initial condition. Owing to the $\pi$-phase difference, there is a repulsive interaction between the two solitons, which is expected to balance the attractive force induced by the focusing nonlocal Kerr nonlinearity, and hence favors the formation of a stable two-soliton molecule~\cite{potential}.

\begin{figure}[ht]
\centering
\includegraphics[width=0.6\linewidth]{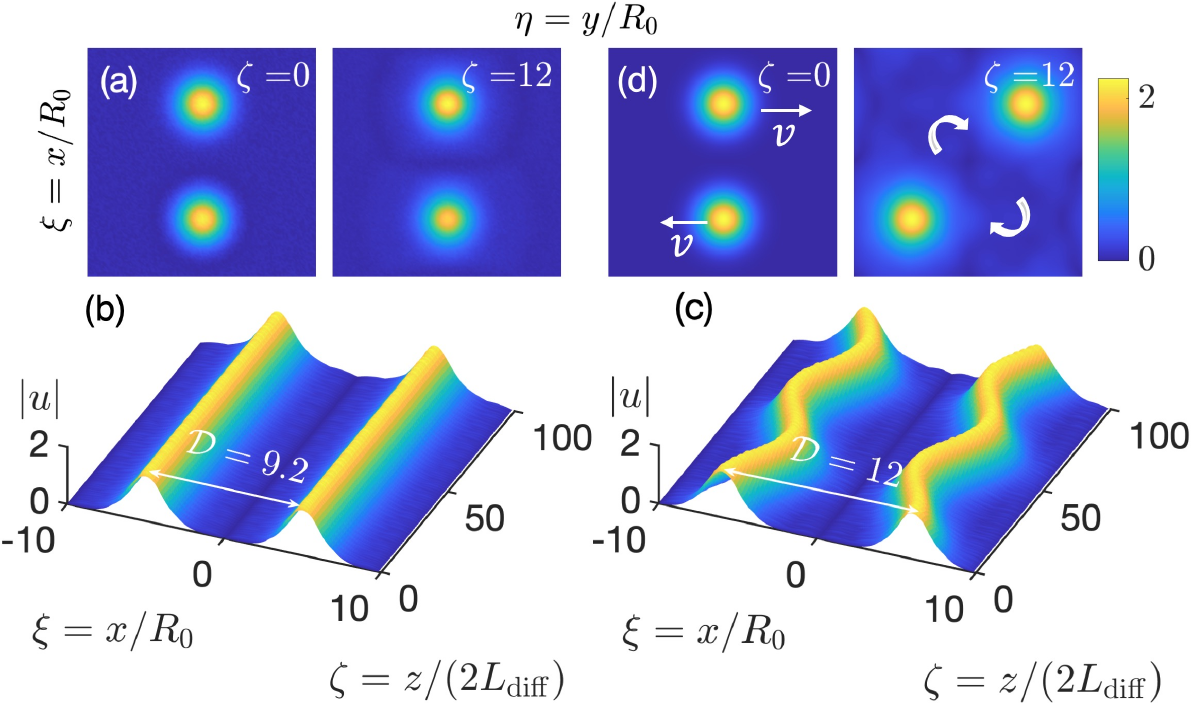}
\caption{\footnotesize   
Propagation of a (2+1)D spatial two-soliton molecule.
(a)~Amplitude profiles $|u|$, at different propagation distances, $\zeta=0$ and$\,12$. When two solitons are
When placed in equilibrium positions, they form stable molecules. 
(b) Same as (a), but for $|u|$ in the cross-section with $\eta=0$.
(c)When the solitons are initially shifted from their equilibrium positions, they perform small oscillations around the equilibrium.
(d) Same as (a), but with an initial velocity $v=\pm0.5$. In this case, the white arrows indicate the direction of rotation of the emerging SM. Figures are adapted from Ref.~\cite{qin2022stable}.
}
\label{fig3}
\end{figure}
Figure~\ref{fig3}(a)
shows the amplitude of a typical two-soliton molecule at $\zeta =0$ and $12$. The initial condition for the simulation are chosen as per ansatz (\ref{ansatz0}) with a small random perturbation introduced by the factor ${\rm Random}=1+\epsilon R(\xi,\eta)$, 
by multiplying the initial configuration. Here $\epsilon \ll 1$ is the amplitude of the perturbation, and $R$ is a random variable uniformly distributed in the interval [$0,1$]. The SM is found to be stable as it relaxes to the self-cleaned form close to the unperturbed one and undergoes no apparent distortion during propagation. The cross-section $\eta=0$ of the SM propagation is shown in Fig. ~\ref{fig3}(b), respectively. However, when the two solitons are slightly shifted from their equilibrium positions,
they perform a small oscillation around the equilibria if no additional perturbations are introduced; see Fig.~\ref{fig3}(c). 
Adding initial velocities $v=\pm 0.5$ to the solitons in Fig.~\ref{fig3}(d), the SM exhibits persistent rotation, keeping its stability in the course of the propagation. As rotation results in an additional centrifugal force acting on each soliton, the size of the rotating SM is larger than that of the non-rotating SM.

\subsection{Light bullets and light vortex bullets}

Light bullets (LBs) are high-D spatiotemporal optical solitons localized in $m$-spatial dimensions and one time axis [$(m+1)$D; $m=1$, 2, 3]. In recent years, the study of LBs has attracted intensive theoretical and experimental interest because of their rich nonlinear physics and promising applications. However, LBs generally suffer from severe spread or collapse instability, which limits their propagation to only a few diffraction lengths. A commonly used approach for overcoming such instability and forming stable LBs is to exploit local and nonlocal optical nonlinearities with different response times. Due to the existence of both local and nonlocal Kerr nonlinearities, Rydberg atomic gas may provide a new and ideal platform for exploring stable (3+1)D LBs~\cite{Bai2019,Dong:22}.

\begin{figure}[ht]
\centering
\includegraphics[width=0.69\linewidth]{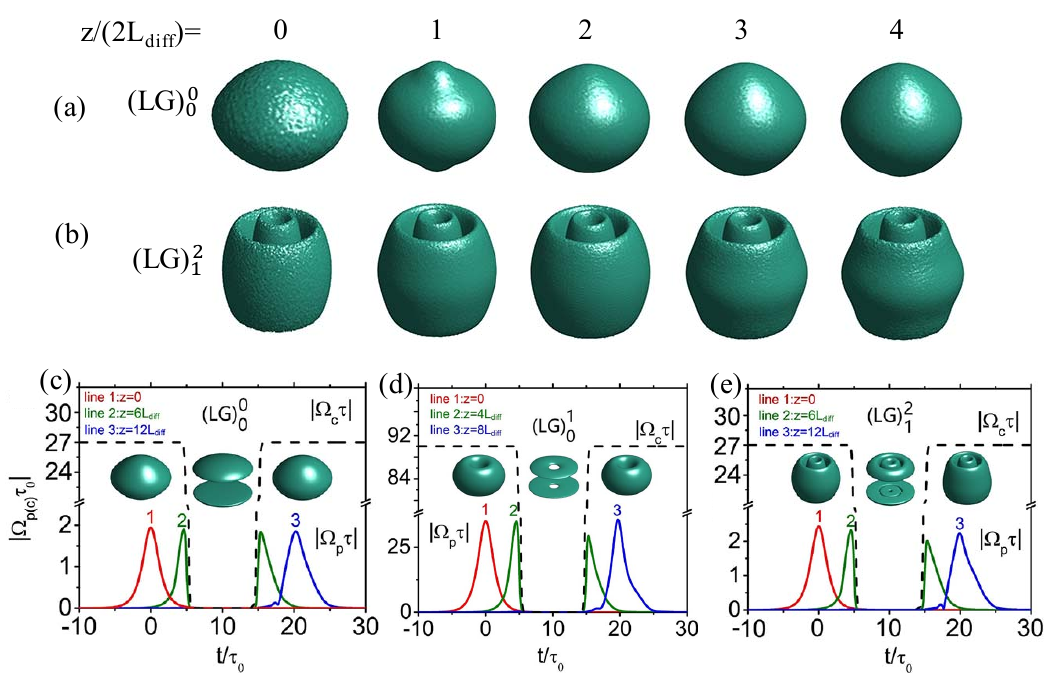}
\caption{\footnotesize   
Propagation of LBs and LVBs in the strongly nonlocal response region. Evolution of $|u|^2$ with a typical LB [panel (a)] and a typical LVB with the (LG)$_1^2$ mode [panel (b)] at different propagation distances.
(c), (d) and (e)~Storage and retrieval of (3+1)D LBs and LVBs in the strong nonlocal response region. (c) Memory of the LB (with (LG)$_0^0$ mode). The black dashed line shows the switching on and off of the control field $|\Omega_c\tau_0|$. Curves 1, 2, and 3 are temporal profiles of the probe pulse $|\Omega_p\tau_0|$, respectively, at $z = 0$ (before storage), $z = 6L_{\rm diff}$  (at the beginning of storage), and $12L_{\rm diff}$ (after storage); the corresponding isosurface plots for $|\Omega_p\tau_0|=0.1$ are also shown. (d) and (e) The same as (c) but for the memory of the LVB with (LG)$_0^1$ and (LG)$_1^2$ modes. Figures are adapted from Ref.~\cite{Bai2019}.
}
\label{Ch2_bullets}
\end{figure}

To find the LB solutions of the (3+1)D NNLSE~(\ref{NNLSE1}), we employ the variational approach using the ansatz:
\begin{align}
 u= A_s\exp \left(-\frac{\xi^2+\eta^2}{w_s^2}\right) {\rm sech}\left(\frac{\sigma}{w_t}\right)\exp\left[-iC_s\frac{\xi^2+\eta^2}{w_s^2}-iC_t\frac{\sigma^2}{2}+i\phi\right].
\end{align}
Here, the parameter $A_s$ is the amplitude, $w_s$ is the transverse beam radius, $w_t$ is the pulse duration, $C_s$ is the wavefront curvature, $C_t$ is the temporal chirp of the probe pulse, and $\phi$ is the phase. All these parameters depend on $\zeta=z/(2L_{\rm diff})$. Moreover, Eq.~(\ref{NNLSE1}) admits stable (3+1)D LBs carrying orbital angular momenta, called light vortex bullets (LVBs). To demonstrate this, we take the ansatz with the form
\begin{align}
 u_{m p}= \frac{C_{m p}}{\sqrt{w_s}}\left[\frac{\sqrt{2} \sqrt{\xi^2+\eta^2}}{w_s}\right]^{|m|} \exp \left(-\frac{\xi^2+\eta^2}{w_s^2}\right)L_p^{|m|}\left[\frac{2\left(\xi^2+\eta^2\right)}{w_s^2}\right] \operatorname{sech}\left[\frac{\sigma}{w_t(s)}\right] e^{i m \varphi},
\end{align}
where $L_p^{|m|}$ is the generalized Laguerre–Gauss (LG) polynomial with $m$ and $p$ as the radial and azimuthal indices, respectively. The stability of the LBs and LVBs can be investigated through linear stability analysis.

The system allows for the generation of stable LBs and LVBs, as shown in Figs.~\ref{Ch2_bullets}(a) and (b). In these figures, the propagation of a typical (3+1)D LB and a typical (3+1)D LVB with (LG)$_1^2$ mode are illustrated for different propagation distances. We observe that the LB and LVB relax to their self-cleaned forms that are quite close to the unperturbed ones when small noises are introduced in their initial conditions, and their shapes undergo no apparent change during propagation.


One of the main advantages of EIT is the possibility of the active manipulation of optical pulses by tuning the system parameters. In particular, optical pulses can be stored and retrieved by switching the control field off and on. In recent years, several studies have been conducted to build light memories using EIT systems~\cite{Maxwell2013,li_quantum_2016}. However, it is generally difficult to realize the memory of high-D nonlinear optical pulses via conventional EIT systems because of their spread or collapse during propagation. Fortunately, the storage and retrieval of stable LBs and LVBs with high efficiency and fidelity are possible in the Rydberg-EIT system.
In particular, we investigate the propagation of LBs and LVBs using a control field that is switched on and off adiabatically, which can be described by the switching function of the form $\Omega_c(t)=\Omega_{c 0}\left\{1-1 / 2 \tanh \left[\left(t-T_{\text {off }}\right) / T_s\right]+1 / 2 \tanh \left[t-T_{\text {on }} / T_s\right]\right\}$, where $T_{\rm off}$ and $T_{\rm on}$ are the times at which the control field is switched off and on, respectively. The duration of the switching time is $T_s$, and the storage time of the probe pulse is approximately given by $T_{\rm on} - T_{\rm off}$.

Figure~\ref{Ch2_bullets}(c) shows the numerical result of the intensity of probe pulse $|U/U_0|^2$ during storage ($z=6L_{\rm diff}$) and retrieval ($z=12L_{\rm diff}$). When the control field $\Omega_c$ is switched on, the (3+1)D LB is created; by switching off the control field, the LB is stored in the atomic medium; then, the LB is retrieved when $\Omega_c$ is switched on again. During storage, the LB information is converted into that of the atomic spin wave (i.e., the coherence $\rho_{13}$). Note that the retrieved LB has nearly the same wave shape as that before storage. The slight deformation of the LB after storage is due to dissipation (dephasing and spontaneous emission) and the small imbalance between the diffraction, dispersion, and nonlinearities in the system. Storage and retrieval of LVBs can also be achieved; see Fig.~\ref{Ch2_bullets}
(d) and (e) the LVB with (LG)$_0^1$ and (LG)$_1^2$ modes.

Following the above investigation, it is shown that stable (3+1)D vector light bullets with ultraslow propagating velocity and extreme low generation power can be realized in a cold Rydberg atomic gas. They can also be actively controlled using a nonuniform magnetic field; in particular, the trajectories of their two polarization components can have significant Stern–Gerlach deflections~\cite{Mu2022OL}.


\subsection{Self-induced transparency solitons}

\begin{figure}[ht]
\centering
\includegraphics[width=0.7\linewidth]{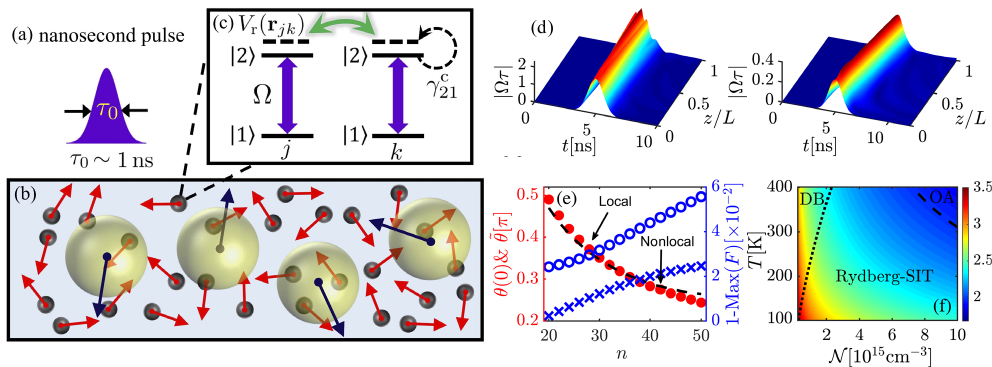}
\caption{\footnotesize   
Light-atom interactions in thermal gases. (a) Nanosecond laser pulses excite atoms from the ground state to the Rydberg states. (b) In a gas of warm atoms, the excitation is affected by thermal motion, Rydberg atom interactions, and inelastic collisions between ground-state (black dots) and Rydberg (yellow balls) atoms. 
(c) Level scheme. The laser (Rabi frequency $\Omega$) resonantly couples the ground state $|1\rangle$ and Rydberg state $|2\rangle$. The latter experiences strong, long-range van der Waals interactions $V_r({\bf r}_{jk})$ and collisional decay (rate $\gamma_{21}^{c}$). 
(d) Rydberg SIT and the optimal area. The pulse is distorted when $\theta(0)=2\pi$ (left panel) and stable when $\theta(0)=0.35\pi$ (right panel), corresponding to the Rydberg SIT.
(e) Optimal area (filled circle) and corresponding fidelity at 1 $\mu$K (star) and 300 K (empty circle). The MF (dashed line) and numerical calculations are in good agreement. (f) Three regimes of nanosecond pulses dominated by either Doppler broadening (DB), Rydberg interactions, or optical absorption (OA). The Rydberg SIT forms in the Rydberg interaction-dominant region. 
Figures are adapted from Ref.~\cite{bai_self-induced_2020}.
}
\label{Ch2_SIT}
\end{figure}

In the conventional EIT approach, ultracold temperatures are critical for maintaining dispersive nonlinearity. As Doppler broadening $\propto\sqrt{T}$, with $T$ being the temperature, large thermal fluctuations at high temperatures can easily smear out the nonlinearity. To overcome this limitation, some experimental and theoretical studies have been reported using short (nanoseconds) and strong (gigahertz Rabi frequencies) lasers to excite high-density, room-temperature (or hot) Rydberg gases confined in glass cells. Moreover, it was found that a short laser pulse can be shaped into a bright soliton, without suffering any absorption or distortion, leading to Rydberg self-induced transparency (SIT) in high-temperature gases.

Specifically, we consider a nanosecond laser pulse propagating along the $z$ axis in a high-density gas (density $\mathcal{N}$), as depicted in Figs.~\ref{Ch2_SIT}(a) and (b). The laser resonantly couples ground state $|1\rangle$ to Rydberg $nP$ state $|2\rangle$ via a single-photon transition [see Fig.~\ref{Ch2_SIT}(c)]. Two Rydberg atoms (located at ${\bf r}_j$ and ${\bf r}_k$) interact via the van der Waals (vdW) interaction  $V({\bf r}_{jk})=-C_6/|r_{jk}|^6$, where ${\bf r}_{jk}={\bf r}_j-{\bf r}_k$ and $C_6 \propto n^{11}$ is the dispersion coefficient. 

In the absence of Rydberg-Rydberg interactions, SIT occurs if the areas of the input pulse $\theta(z)=\int_{-\infty}^{\infty}\Omega(z)dt=\Omega_s \tau \pi$ are multiples of $2\pi$, that is, $\Omega_s \tau$ is an even number, governed by the area theorem~\cite{lamb_RevModPhys.43.99}. This nonlinear effect is solely rooted in high light intensities, which reshape the pulse into a stable, bright soliton, i.e., without absorption or distortion. 

In the presence of a strong Rydberg-Rydberg interaction, the pulse profile is distorted when the input area $\theta(0)=2\pi$ [left panel of Fig. ~\ref{Ch2_SIT}(d)]. However, its shape is preserved for a smaller input area, e.g., $\theta(0)=0.35\pi$ for $n=30$ [right panel of Fig.~\ref{Ch2_SIT}(d)], which gives rise to Rydberg SIT. Similar to conventional SIT, the formation of the Rydberg SIT can be understood by analyzing the atomic dynamics~\cite{lamb_RevModPhys.43.99}, which are independent of $z$ as the nanosecond pulse is translated in the medium.
It shows that the optimal area decreases monotonically with increasing $n$, while the corresponding fidelity is high [Fig.~\ref{Ch2_SIT}(e)].

Depending on the ratio $(kv_T+u)/\gamma_{21}^c$, there are three different coherence regimes. Here, $\gamma_{21}^c\propto\sqrt{T}$ is the decay rate due to the inelastic collision, $v_T=\sqrt{2k_BT/M}$
is the thermal velocity ($M$ is the mass of the Rydberg atoms), and $u=2{\cal N}^{1/3} \int_0^\infty dzV(z)$ is the effective Rydberg-Rydberg interaction (${\cal N}$ is the density of Rydberg atoms).
Fixing $T$, a Doppler broadening dominant region appears at low densities when
$kv_T>u\gg\gamma_{21}^{c}$,
as shown in Fig.~\ref{Ch2_SIT}(f). For sufficiently high
densities [dotted lines in Fig. ~\ref{Ch2_SIT}(f) with $10kv_T=u$] Rydberg-Rydberg interactions overtake the other two effects, that is, $u > kv_T\gg \gamma_{21}^c$. This is the most interesting region where the Rydberg SIT can form. At further increasing densities (dashed line, $ kv_T+u=100 \gamma_{21}^c$), the collisional decay starts to kick in and causes losses. The overall decay also depends on the propagation distance.

\subsection{Other topics for the focusing Rydberg atomic gas}

Beyond standard EIT-based schemes, Rydberg atomic gases provide a versatilenonequilibriumploring advanced optical and many-body phenomena in quantum physics. Here, we highlight several emerging directions that leverage strong, long-range Rydberg interactions, organized from coherent nonlinear wave mixing to complex nonequilibrium dynamics and structured light fields.

Four-wave mixing (FWM) is a fundamental nonlinear process that combines three optical waves to generate the fourth. In Rydberg thermal vapors, both theoretical and experimental studies have shown that strong Rydberg–Rydberg interactions can dramatically enhance the FWM efficiency~\cite{Brekke2008PRA, Pfau2012PRA}. For example, experiments in rubidium Rydberg vapors have demonstrated FWM efficiencies that are orders of magnitude larger than those in conventional atomic systems, owing to the giant nonlocal Kerr nonlinearity that modifies the phase-matching conditions and enhances the nonlinear coupling.  Building on this, higher-order processes, such as six-wave mixing (SWM), have been realized, where six fields interact to produce new frequency components. A notable application is coherent microwave-to-optical conversion in cold $^{87}$Rb atoms, achieved via SWM: a microwave field couples two Rydberg states, yielding a conversion efficiency of $\sim0.3\%$ over a bandwidth exceeding 4 MHz~\cite{LiWH2018PRL}. Subsequently, the efficiency was improved to $\sim5\%$ in the linear regime, with a realistic pathway outlined to surpass $60\%$~\cite{LiWH2019PRA}. 
The key enabling factors are the long Rydberg lifetimes and the strong dipole coupling to millimeter-wave and microwave radiation. 
However, several challenges remain, including the optimization of phase-matching conditions, the suppression of losses arising from dephasing and spontaneous emission from intermediate states, and the careful tuning of atomic density and interaction strengths to avoid detrimental collisional effects. 

Because Rydberg-EIT systems allow active and precise control, a standing-wave control field can be used to create an electromagnetically induced grating (EIG)~\cite{hang2019PRA, gao2022PRA, qian2019PRA}. This enables the study of the nonlocal nonlinear diffraction of the probe field. In the far-field regime, tuning the EIG depth via the system parameters leads to either Raman–Nath or Bragg diffraction~\cite{hang2019PRA, gao2022PRA}. In the near-field regime, nonlocal nonlinear Talbot carpets—periodic self-imaging patterns—have been observed and analyzed~\cite{hang2019PRA}.

The interplay between Rydberg interactions and EIT offers an ideal testbed for investigating nonequilibrium many-body phenomena. Recent experiments have explored non-Hermitian physics and nonlinear dynamics in cold Rydberg gases~\cite{Youli2026PRL}. For instance, interaction-induced second-order exceptional points (EPs) have been observed, along with hysteresis trajectories that reveal the breaking of charge-conjugation parity symmetry~\cite{Ding2024}. The application of a microwave field can drive a dynamical topological phase transition, flipping the winding number of the hysteresis loop from $-1$ to $+1$~\cite{zhang2024dynamical}. Third-order EPs have also been reported, with the hysteresis loop area characterized as a function of the scan time and optical density. In a thermal cesium vapor, a transition from bistability to multistability was discovered, including a hidden critical point and up to seven spectral jumps corresponding to higher-order symmetry breaking~\cite{ma2024folded}. 

Complementing these results, Wu {\it et. al} developed superatom models and achieved giant cross-Kerr nonlinearity with $\pi$-phase shifts for all-optical deflection~\cite{Wu2015}. Wu {\it et. al} used thermal Rydberg gases to study nonequilibrium many-body physics and precision measurements, experimentally observing dissipative time crystals, where continuous driving yields persistent, noise-robust photon transmission oscillations~\cite{you2024}. Su {\it et. al} made theoretical contributions to quantum information processing, designing high-fidelity geometric quantum gates resilient to operational errors via Rydberg blockade and antiblockade effects~\cite{Su2024, Su20242}.

Finally, the Rydberg-EIT system supports highly nonlocal nonlinear optical X-waves, which exhibit low propagation loss, ultraslow velocity, and ultralow generation power. Their stability domain can be significantly enlarged by increasing the degree of nonlocality of the Kerr nonlinearity, and their trajectory can be manipulated by an external magnetic field~\cite{hang2020PRA_Xwave}. Furthermore, stable optical Ferris wheel (OFW) solitons have been predicted in a nonlocal Rydberg-EIT medium~\cite{Qiu2023OL, Hamedi2021OL}. Simulations show that soliton fidelity remains above $0.96$ even after propagation exceeding $160$ diffraction lengths, with higher-order OFW solitons of arbitrary winding numbers also being discussed~\cite{Qiu2023OL}.

\section{Defocusing nonlocal nonlinearity}

\subsection{Pattern formation}

A prominent manifestation of self-defocusing nonlinearity in Rydberg media is the spontaneous formation of spatially ordered optical patterns through transverse modulational instability (MI)~\cite{Segur2007,Zakharov2009,Zakharov2013}. While MI in local nonlinear optics is typically associated with self-focusing nonlinearity, the situation is fundamentally different in nonlocal systems. In Rydberg-EIT media, the nonlocal Kerr response arising from long-range atomic interactions allows MI to occur even when the nonlinearity is repulsive.

\begin{figure}
\centering\includegraphics[width=.9\columnwidth]{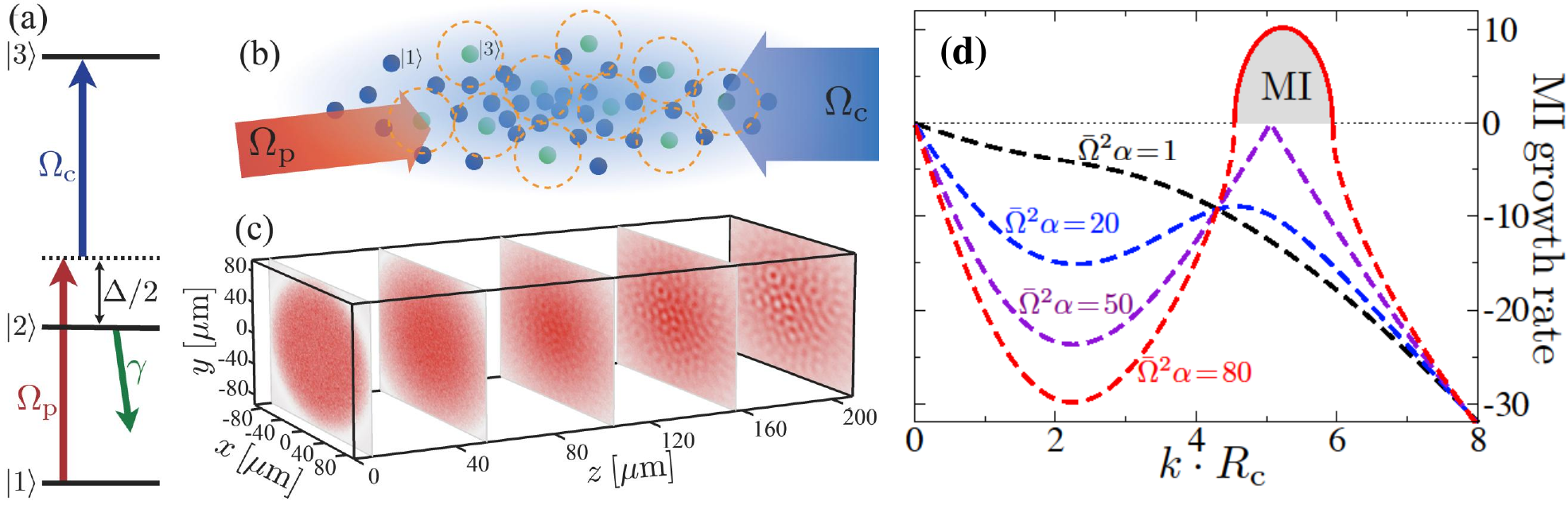}
\caption{\footnotesize (a) Three-level scheme for Rydberg-EIT.
(b) Strong van der Waals interactions between atoms in the Rydberg state $|3\rangle$ inhibit multiple Rydberg excitations within a blockade sphere (red dashed circle), giving rise to a strongly nonlinear optical response of the medium. The resulting nonlinear beam propagation leads to modulation instabilities, as shown in (c) for a Rubidium $70S_{1/2}$ Rydberg gas.
(d) Growth rate $\Gamma_{\rm MI}$ of intensity modulations with wavenumber $k$ for \emph{defocusing} nonlinearities of different strengths ${\bar \Omega}^2\alpha$. Figures are adapted from Ref.~\cite{sevinccli2011PRL}
}
\label{shi-fig1}
\end{figure}

In ladder-type Rydberg-EIT configurations with repulsive van der Waals interactions as shown in Fig.~\ref{shi-fig1}(a), it has shown that a homogeneous plane-wave probe field becomes unstable against transverse perturbations within a finite annular region of momentum space. The nonlinear development of this instability leads to spontaneous symmetry breaking and the emergence of stationary optical-lattice structures. Among the possible solutions, hexagonal lattice patterns are energetically favored and dynamically stable~\cite{sevinccli2011PRL}. These hexagonal patterns arise from the resonant coupling of three transverse modes with equal wave numbers, reflecting the rotational symmetry of the MI gain spectrum in isotropic nonlocal media [see Fig.~\ref{shi-fig1}(d)].  Importantly, pattern formation occurs without any externally imposed transverse potential, demonstrating genuine optical self-organization mediated solely by Rydberg-induced nonlocal nonlinearity.

\begin{figure}
\centering\includegraphics[width=.6\columnwidth]{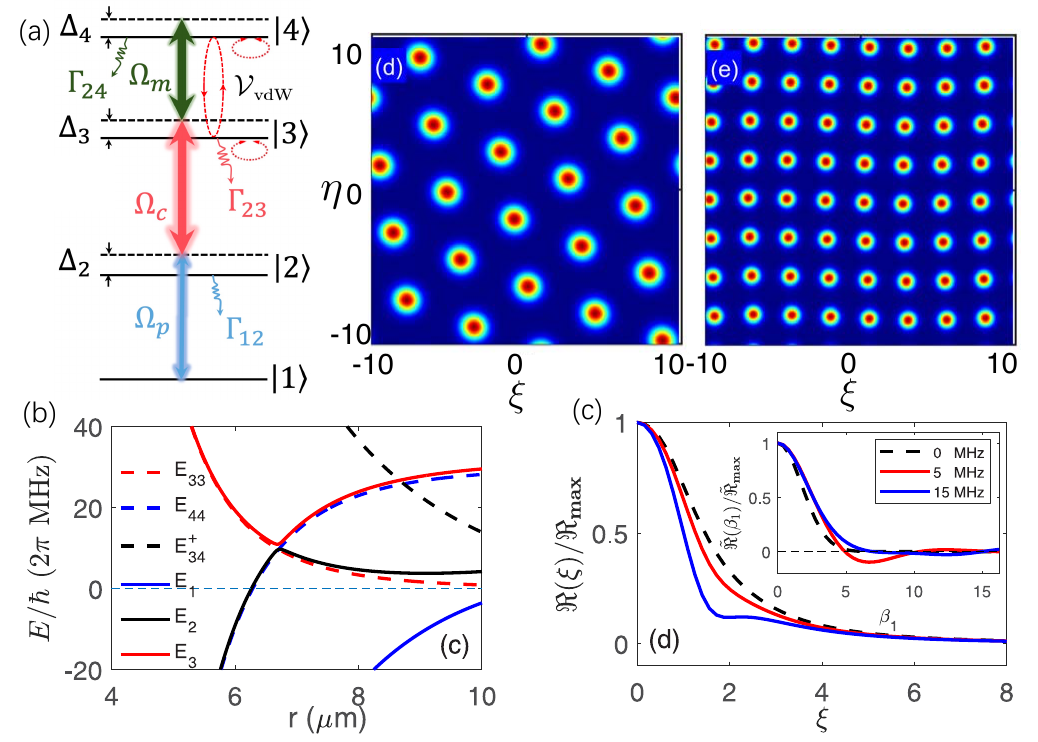}
\caption{\footnotesize (a)~Ladder-type four-level atomic configuration for realizing the microwave-dressed Rydberg-EIT. 
Both $|3\rangle$ and $|4\rangle$ are highly excited Rydberg states.
Two Rydberg atoms located at positions ${\bf r}$ and ${\bf r}'$ interact through the van der Waals potential $\hbar {\mathcal V}_{\rm vdw}^{l}({\bf r}^{\prime}-{\bf r})$ ($l=s,\,d,\,e$).
(b)~Potential-energy curves $E_{1}$ (solid blue line), $E_2$ (solid black line), and $E_3$ (solid red line) of two Rydberg atoms
as functions of the interatomic distance $\rm{r}= |{\bf r'-r}|$ for $\Omega_m=10~$MHz and $\Delta=13.98$ MHz; $E_{33}$ (dashed blue line), $E_{44}$ (dashed black line), and $E_{34}^{+}$ (dashed red line) are for the case without a microwave field.
(c)~Normalized response function $\Re/\Re_{\rm max}$ as a function of the dimensionless coordinate $\xi$.
(d) and (e)~Hexagonal and square lattices of the normalized probe-field amplitude $|u|^2$ as a function of $\xi$ and $\eta$ for $\Omega_m=10$~MHz and $I_{\rm eff}=15$. Figures are adapted from Ref.~\cite{Shi2020}.
}
\label{shi-fig2}
\end{figure}

The controllability of optical self-organized patterns can be significantly enhanced by introducing microwave dressing~\cite{Petrosyan2014} between two highly excited Rydberg states. In microwave-dressed Rydberg-EIT systems, as illustrated in Fig.~\ref{shi-fig2}(a), the strength, range, and effective sign of the nonlocal Kerr nonlinearity can be actively manipulated while maintaining low optical absorption, as shown in Figs.~\ref{shi-fig2}(b) and (c). By tuning the microwave Rabi frequency, the MI landscape of the probe field is substantially modified, allowing multiple competing stationary patterns to emerge.

As the microwave field strength increases, the system undergoes structural phase transitions between distinct self-organized optical states~\cite{Shi2020}. In addition to the hexagonal lattice pattern observed in the absence of microwave dressing, different types of square lattice patterns can be stabilized, as shown in Figs.~\ref{shi-fig2}(d) and (e). These phase transitions are controlled by experimentally accessible parameters, including microwave field strength, effective probe intensity, and degree of nonlocality. This behavior represents a rare example of actively tunable phase transitions of spatially ordered light in conservative nonlinear optical systems.

\begin{figure*}
  \centering
  \includegraphics[width=0.9\textwidth]{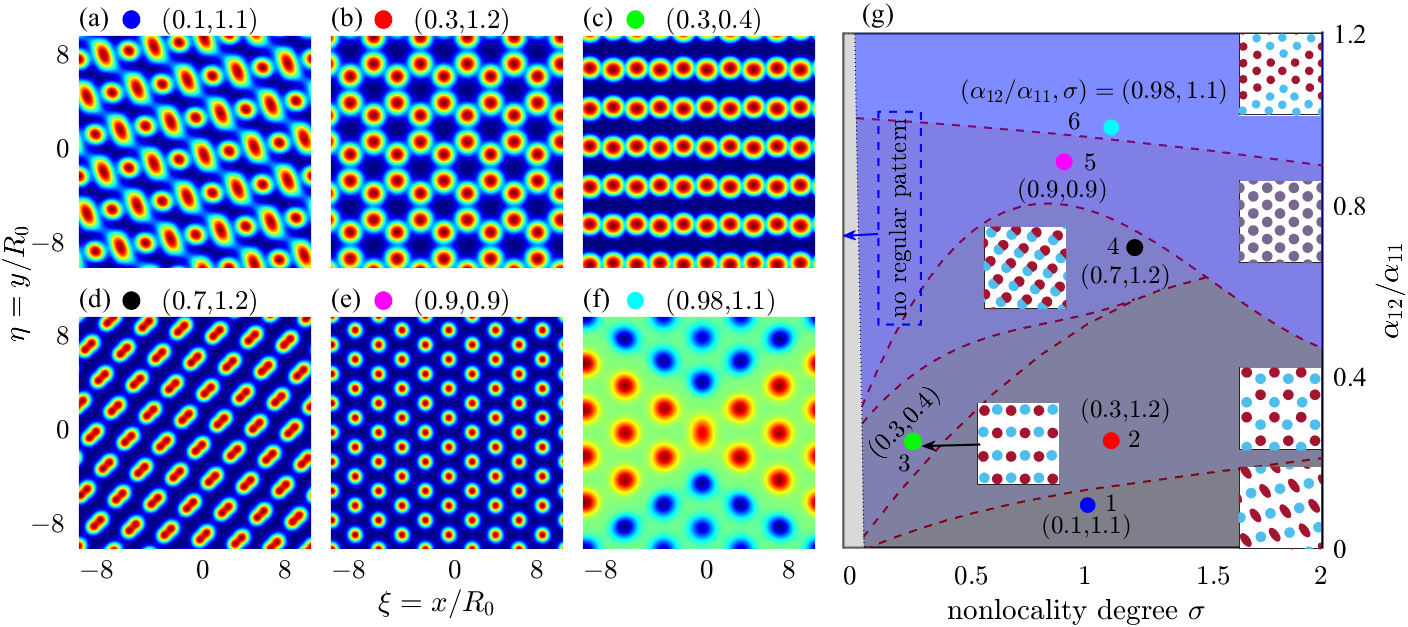}
\caption{\footnotesize Pattern formation of the two-component probe field controlled by the ratio of nonlinearity $\alpha_{12}/\alpha_{11}$ and the nonlocality degree $\sigma$, by taking the non-dimensional normalized probe-field intensity $|v|^2=|v_1|^2+|v_2|^2$ as a function of the non-dimensional coordinates $\xi = x/R_0$ and $\eta = y/R_0$ (with  $\alpha_{11}=550$  and $\rho_{22}^{(0)}/\rho_{11}^{(0)}=1$) for different $\alpha_{12}/\alpha_{11}$  and $\sigma$:
(a) $(\alpha_{12}/\alpha_{11}, \sigma)= (0.1, 1.1)$;
(b) $(\alpha_{12}/\alpha_{11}, \sigma)= (0.3, 1.2)$;
(c) $(\alpha_{12}/\alpha_{11}, \sigma)= (0.3, 0.4)$;
(d) $(\alpha_{12}/\alpha_{11}, \sigma)= (0.7, 1.2)$;
(e) $(\alpha_{12}/\alpha_{11}, \sigma)= (0.9, 0.9)$;
(f) $(\alpha_{12}/\alpha_{11}, \sigma)= (0.98, 1.1)$.
(g)~Phase diagram for the optical pattern formation, in which different domains (phases) are obtained by using different values of ($\alpha_{12}/\alpha_{11}$,\,$\sigma$) corresponding respectively to those used in (a)-(f). Figures are adapted from Ref.~\cite{Shi2021}.
}
\label{shi-fig3}
\end{figure*}

Further richness in pattern formation arises in systems supporting multicomponent probe fields under double Rydberg-EIT conditions~\cite{Shi2021}, which is given in Fig.~\ref{shi-fig3}. In such configurations, both the self-Kerr and cross-Kerr nonlinearities are nonlocal and of comparable magnitude, resulting in CNNLSE for different probe field components. A detailed MI analysis and nonlinear evolution revealed a wide variety of stationary optical patterns, including hexagonal, square, stripe, and mixed-symmetry lattice structures.

\begin{figure}
\centering\includegraphics[width=0.8\linewidth]{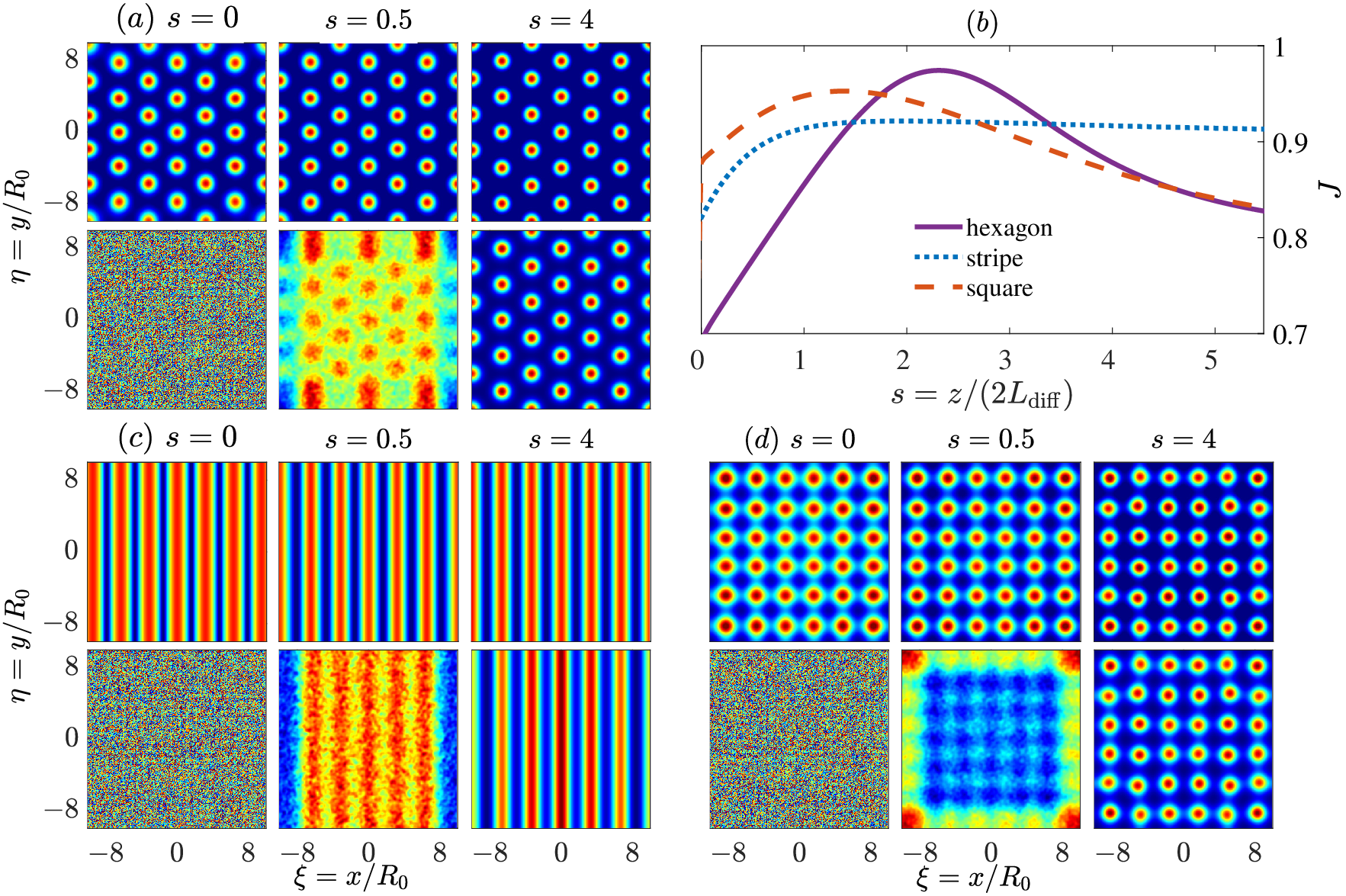}
\caption{\footnotesize
Cloning of different optical self-organized structures. Normalized probe field 1 (the cloned light; top) and cloned light (probe field 2; bottom) intensity distributions as functions of $\xi$ and $\eta$ at different propagation distances for (a) hexagonal, (c) stripe, and (d) square structures, with $s=z/(2L_{\rm diff})=0,\,0.5,\,\,4$.  
(b) Fidelity $J$ of cloned structures as a function of propagation distance $s=z/(2L_{\rm diff})$. The purple solid line, blue dotted line, and red dashed line represent the fidelity curves for the hexagonal, stripe, and square structures, respectively. Figures are adapted from Ref.~\cite{Shi2021OL}.
}\label{shi-fig4}
\end{figure}

Optical phase separation is an important phenomenon in multicomponent systems. When the ratio of multicomponentself-Kerr nonlinearities exceeds a critical threshold, the system undergoes a transition from spatially mixed to spatially separated configurations of different probe field components, as shown in  Fig.~\ref{shi-fig3}(f). This behavior closely parallels phase separation in multicomponent quantum fluids, highlighting the deep connections between optical self-organization in Rydberg media and many-body physics.

Beyond spontaneous pattern formation, the controlled selection and stabilization of optical patterns can be achieved by introducing a weak spatial modulation of the control laser field~\cite{Shi2021OL}. Such modulation slightly reshapes the MI gain spectrum and lifts the degeneracy among unstable modes, enabling the deterministic selection of stripe or square patterns that are otherwise unstable in homogeneous configurations. The resulting optical structures remain in a self-organized state rather than a trivial imprint of the external modulation, as their spatial periods and symmetries differ from those of the modulated control field.

Nonlocal cross-Kerr nonlinearities in double Rydberg-EIT systems also enable the high-fidelity cloning of optical patterns~\cite{Shi2021OL}. Periodic structures such as hexagonal, stripe, and square lattices formed in one probe field can be transferred to another initially homogeneous probe field while preserving spatial symmetry in Fig.~\ref{shi-fig4}. This pattern cloning process is robust during propagation and suggests potential applications in all-optical pattern transfer, imaging, and information processing.

\subsection{Dispersion Shock waves}

Nonlinear hydrodynamic flows are found in different media~\cite{whitham1999linear,ablowitz2011nonlinear}, such as 
ultracold quantum gases~\cite{PerezPRL2004,kamchatnov2004PRA,hoefer2006PRA,joseph2011PRL,chang2008PRL,Simmons2023PRA}, and optical media~\cite{el2007PRA,rothenberg1989PRL,wan2007NP,jia2007PRL,ghofraniha2007PRL,Kartashov2013OL,gentilini2015PRA,marcucci2019APX,xu2017PRL,WetzelPRL2016,ContiPRA2010,marcucci2020aPRL}. In defocusing nonlinear media, an initially smooth wave steepens when propagation, eventually reaching a point of gradient catastrophe~\cite{kamchatnov2000book}, that lead
to the formation of shock waves~\cite{courant1944book,kamchatnov2000book,kleine2001book,zel2002book,ghofraniha2012PRL,Hang2023PRA,mossman2018NC,isoard2019PRA,isoard2020EPL,Kamchatnov2020chaos,el2016PRS,Simmons2020PRL,smoller2012book,marcucci2019NC,bienaime2021PRL}.

In order to acquire an understanding of shock wave behaviors, it is helpful to express the NNLSE~(\ref{NNLSE1}) in a hydrodynamic form, and hence treat the light field as a classical fluid. This can be done by using the Madelung transformation
$u(\xi,\zeta)=\sqrt{\rho(\xi,\zeta)}e^{i\phi(\xi,\zeta)}$, which results in two Euler-like fluid equations
\begin{subequations}\label{Euler1}
\begin{eqnarray}
&&\frac{\partial \rho}{\partial \zeta}+\frac{\partial}{\partial \xi}(\rho v)=0,\label{Euler1a}\\
&&\frac{\partial v}{\partial \zeta}+\frac{\partial}{\partial \xi}\left[\frac{1}{2}v^2-{\cal G}\int d\xi' g(\xi'-\xi)\rho(\xi',\zeta) -\frac{1}{2\sqrt{\rho}}\frac{\partial^2 \sqrt{\rho}}{\partial \xi^2} \right]=0\label{Euler1b},
\end{eqnarray}\end{subequations}
where $\rho=|u|^2$ is the light intensity, $v=\partial\phi/\partial \xi$ is the flow velocity of the light fluid, and $\cG$ characterizes the strength of the nonlocal nonlinearity.

In the weak nonlocality regime, we let $g(\xi'-\xi)\rightarrow\delta(\xi'-\xi)$ and omit the quantum pressure~\cite{Hang2023PRA,Qin2024PRA}, arriving at the celebrated shallow-water-like equations for the light fluid of the probe field
\begin{subequations}\label{shallow_water}
\begin{eqnarray}
&&\frac{\partial \rho}{\partial \zeta}+\frac{\partial}{\partial \xi}(\rho v)=0,\\
&&\frac{\partial v}{\partial \zeta}+\frac{\partial}{\partial \xi}\left(\frac{1}{2}v^2+{\cal G}\rho\right)=0.
\end{eqnarray}\end{subequations}
It is convenient to cast these equations into the diagonal Riemann form
\begin{subequations}\label{Riemann_form}
\begin{eqnarray}
&& \frac{\prt r_+}{\prt\zeta}+\frac12(3r_++r_-)\frac{\prt r_+}{\prt\xi}=0,\\
&& \frac{\prt r_-}{\prt \zeta}+\frac12(r_++3r_-)\frac{\prt r_-}{\prt\xi}=0,
\end{eqnarray}\end{subequations}
where the Riemann invariants are given by
$r_{\pm}={v}/2\pm\sqrt{{\cal G}\rho}$. As long as $r_+$ and $r_-$ are found, the light fluid intensity and flow velocity are determined by $\rho=(r_+-r_-)^2/(4\cG)$ and $v=r_++r_-$.



To be concrete, we assume that the initial light fluid intensity and the flow velocity respectively have the form
$\rho(\xi,0)=\rho_b+\rho_h\,e^{-\xi^2/w_h^2}$ and $v(\xi,0)=0$, which can be easily prepared in a real experiment. Here, $\rho_h$ and $w_h$ characterize the peak intensity and width of a Gaussian hump, respectively, added to the uniform background. From Eqs.~(\ref{Riemann_form}) and the initial conditions, 
the intensity can be expressed implicitly with the solution
\begin{equation}\label{Hopf_solution2}
\rho(\xi,\zeta)=\rho_b+\rho_h\,e^{-[\xi-\sqrt{{\cal G}}(3\sqrt{\rho}-2\sqrt{\rho_b})\zeta]^2/w_h^2}.
\end{equation}

Since the flow velocity $v$ depends on the light intensity $\rho$, the hump exhibits indeed a self-steepening in the direction of propagation, resulting in a gradient catastrophe $\partial_\xi \rho(\xi,\zeta)=-\infty$ at
a certain distance $\zeta=\zeta_{\rm wb}$. This gradient catastrophe leads to wave breaking, followed by shock wave formation. The distance $\zeta_{\rm wb}$ is called the wave breaking
distance, which can be determined from the conditions
$\partial \xi(\rho)/\partial \rho=0$ and $\partial^2 \xi(\rho)/\partial \rho^2=0$, yielding the solution
\begin{equation}\label{wave_breaking1}
\zeta_{\rm wb}=\frac{w_h}{3\sqrt{{\cal G}}}\frac{\sqrt{\rho_s+\rho_b}}{\rho_s-\rho_b}.
\end{equation}
Here $\rho_s$ is the light fluid intensity corresponding to $\partial_\xi \rho_s \rightarrow-\infty$, which can be obtained from the equation $\ln[\rho_h/(\rho_s-\rho_b)]=\rho_s/(\rho_s+\rho_b)$. 

\begin{figure}
\centering\includegraphics[width=0.8\linewidth]{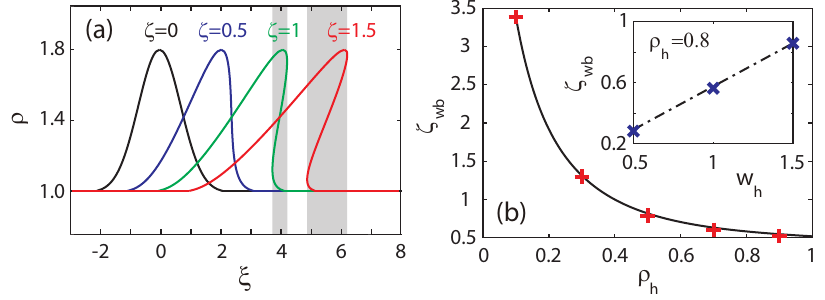}
\caption{\footnotesize
(a)~Light intensity profile $\rho$ as a function of $\xi$ for different distances $\zeta=0$, 0.5, 1, and 1.5. The hump exhibited obvious self-steepening in the direction of propagation. Wave breaking occurs at $\zeta\approx0.7$. (b) Wave breaking distance $\zeta_{\rm wb}$ as a function of the hump's peak intensity $\rho_h$. Inset: The wave breaking distance $\zeta_{\rm wb}$ as a function of
the width of the hump $w_h$. A good agreement is obtained between the analytical prediction (denoted by solid black and dash-dotted blue lines) and the numerical calculation (denoted by symbols ``+'' and ``x'', respectively). Figures are adapted from Ref.~\cite{Hang2023PRA}.
}
\label{shock_BP}
\end{figure}

Figure~\ref{shock_BP}(a) shows the result of the light intensity profile $\rho$ as a function of $\xi$ at different propagation distances $\zeta$. An obvious self-steepening of the hump occurred in the direction of propagation, resulting in wave breaking at a certain distance.
Figure~\ref{shock_BP}(b) is the wave-breaking distance $\zeta_{\rm wb}$ as a function of the hump's peak intensity $\rho_h$ and the hump's width $w_h$ (the inset of the figure).
Thus, there is a good agreement between analytical predictions (denoted by lines) and numerical calculations (denoted by symbols), which means that the diffraction plays indeed no significant effect on the occurrence of the wave breaking.

\begin{figure}
\centering\includegraphics[width=0.8\linewidth]{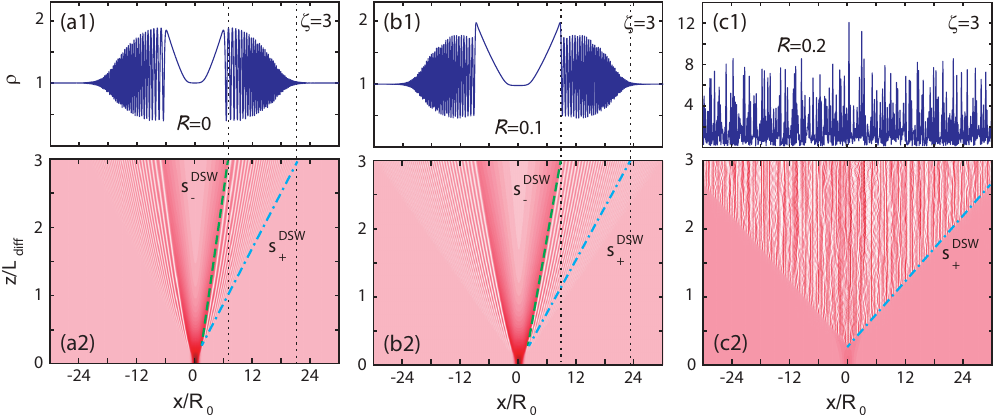}
\caption{\footnotesize
Numerical results of the formation and propagation of the DSW. Snapshots of the intensity profile $\rho=|u|^2$ at $\zeta\equiv z/L_{\rm diff}=3$ (upper panels) and corresponding level plots of the propagation from $\zeta=0$ to 3 (lower panels), for the intensity diffraction parameter ${\cal R}=0$ [(a1) and (a2)], 0.1 [(b1) and (b2)], and 0.2 [(c1) and (c2)], respectively. 
The DSW is stable in the first (${\cal R}=0$) and second (${\cal R}=0.1$) columns, for $c_L\approx2.7<c_{cr}\rightarrow\infty$ and $c_L\approx2.9<c_{cr}\approx3.162$, respectively. However, it is unstable in the third column (${\cal R}=0.2$) for $c_L>c_{cr}\approx2.236$. Figures are adapted from Ref.~\cite{Hang2023PRA}.
}
\label{shockwave_Propagation}
\end{figure}

After wave breaking, a DSW forms owing to the interplay between dispersion (given by the “quantum” pressure) and nonlocal nonlinearity. To demonstrate this, we carry out numerical simulations 
by taking different values of the intensity diffraction parameter ${\cal R}$~\cite{Hang2023PRA}. 
Shown in Fig.~\ref{shockwave_Propagation}(a1) and Fig.~\ref{shockwave_Propagation}(a2)
%
%
are respectively the probe intensity $\rho=|u|^2$ at $\zeta=3$ and its propagation result from $\zeta=0$ to 3 for ${\cal R}=0$. Fig.~\ref{shockwave_Propagation}(b1) and Fig.~\ref{shockwave_Propagation}(b2) show the same results as in Fig.~\ref{shockwave_Propagation}(a1) and Fig.~\ref{shockwave_Propagation}(a2) but for ${\cal R}=0.1$. From the first and second columns, we see clearly that in both cases, the DSWs are quite stable during propagation, which is due to the fact that $c_L<c_{cr}$ (here $c_L$ is the local sound speed of the left (soliton) edge; $c_{cr}=(1/2)\sqrt{\cG/\cR}$ is a critical value of the local sound speed). 
The left-edge local sound speeds in Fig.~\ref{shockwave_Propagation}(a1) and Fig.~\ref{shockwave_Propagation}(b1) are respectively $c_L\approx2.7$ and $c_L\approx 2.9$.
Since $c_{cr}\rightarrow\infty$ ($c_{cr}\approx3.162$) for ${\cal R}=0$ (${\cal R}=0.1$), one has $c_L<c_{cr}$ in both situations. 

The case with a larger intensity diffraction parameter, ${\cal R}=0.2$, is also calculated, with the results presented in Fig.~\ref{shockwave_Propagation}(c1) and Fig.~\ref{shockwave_Propagation}(c2). In this case, however, the DSW becomes unstable, and instability emerges at the small-amplitude edge of the DSW. The instability is caused by the left-edge local sound speed, $c_{L}$, being larger than the critical value $c_{cr}$, that is $c_L>c_{cr}$. Therefore, such an instability originates from the MI of sound waves. 

\subsection{Dissipation effect of Shock waves}

Dissipation plays an important role in the study of Rydberg systems~\cite{Yan:20,hao_observation_2021,dingErgodicityBreakingRydberg2024}. In cold atom gases, dissipation can be induced and controlled~\cite{hang2018PRA,Bai2019}. This opens new opportunities for exploring shock waves in the interplay between nonlocal nonlinearity and controllable dissipation, which is otherwise difficult to achieve in other systems.

\begin{figure}[t]
\centering
\includegraphics[width=0.52\linewidth]{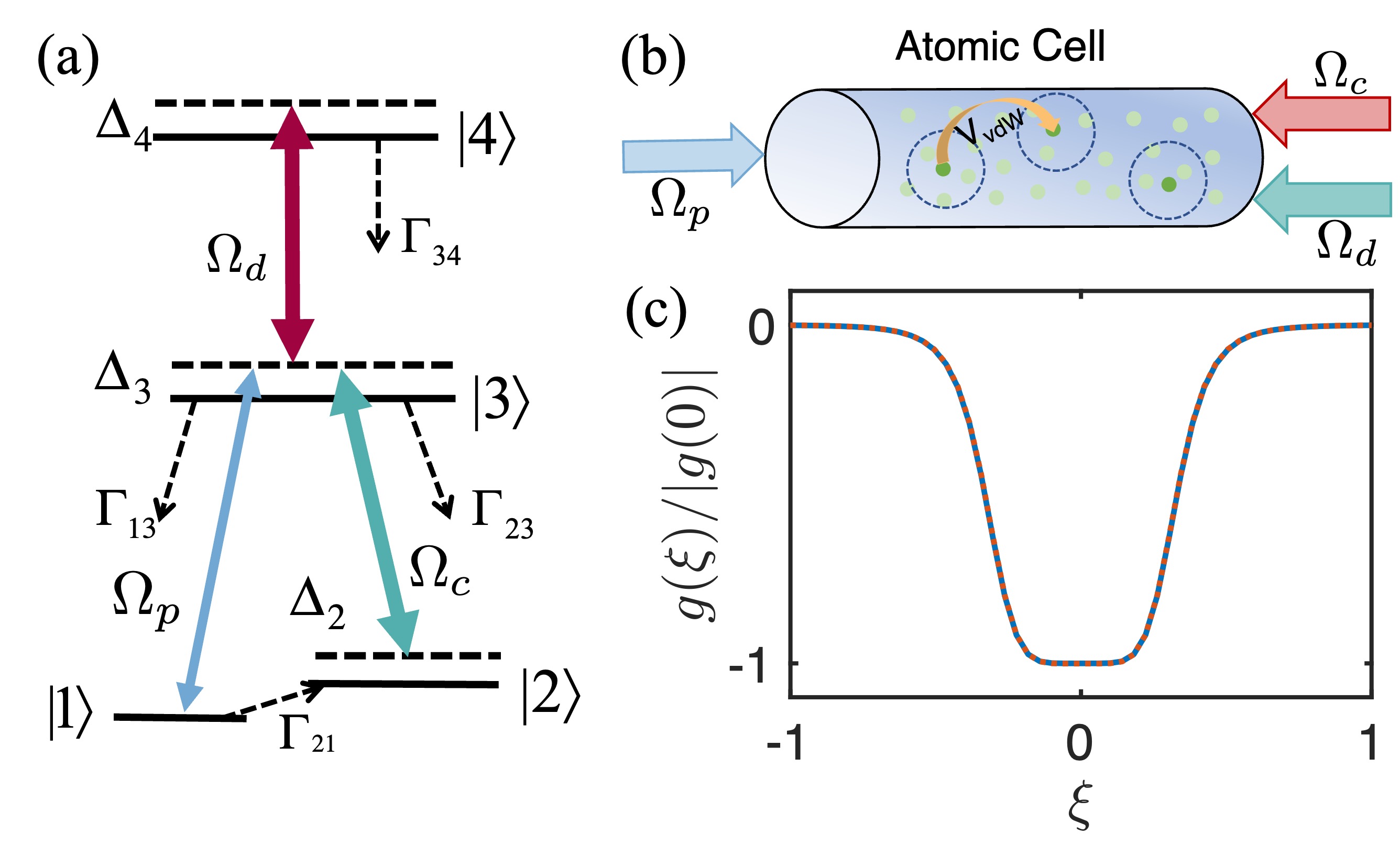}
\caption{\footnotesize
(a)~Inverted Y-type level diagram. To achieve gain, we apply incoherent pumping $\Gamma_{21}$ between states $|1\rangle$ and $|2\rangle$.
(b) Blue, red, and cyan arrows indicate the propagation directions of the probe, control, and dressed fields, respectively. The strong Rydberg atom interaction gives rise to a blockade radius (dashed circle) around the Rydberg atoms. The blockade radius can be tuned using an excitation laser. (c) Response function $g(\xi)$. The numerically obtained $g(\xi)$ (blue solid line) agrees with the analytical approximation~(dashed red line). Figures are adapted from Ref.~\cite{Qin2024PRA}.
} 
\label{fig_model}
\end{figure}

In an inverted Y-type four-level configuration [see Fig.~\ref{fig_model}(a)], the incoherent decay of state $\lvert 3\rangle$ causes loss of the probe field. To generate a gain, incoherent pumping (with pumping rate $\Gamma_{21}$) is used to pump atoms from $|1\rangle$ to $|2\rangle$. Driven by the control laser $\Omega_c$ a small number of atoms are populated in state $|3\rangle$ which provides a gain effect~\cite{hang2018PRA,hang2019PRA}.
Note that in the inverted Y-shaped excitation scheme shown in Fig.~\ref{fig_model}(a), the transition $|1\rangle \rightarrow|2\rangle \rightarrow|3\rangle$ forms a $\Lambda$-type EIT, while $|1\rangle \rightarrow|3\rangle \rightarrow|4\rangle$ forms a ladder type  Rydberg-EIT. The interplay between the two paths can be controlled by external lasers, resulting in dissipative, nonlocal nonlinear interactions. 
This allows to derive a (2+1)D NNLS equation of the probe field~\cite{hang2018PRA,hang2019PRA,shi2023chaos} with dissipative potential, 
\begin{equation}\label{DE1}
i\frac{\partial u}{\partial \zeta}+\frac{1}{2}\frac{\partial^2 u}{\partial \xi^2}-\mathcal{V}u
+g_0\int d\xi'g(\xi',\xi)|u(\xi',\zeta)|^2u=0.
\end{equation}
For convenience,
we have assumed that the strong control field is a plane wave field, resulting in a homogeneous complex potential $\mathcal{V}$. The real part of the complex potential is associated with the refractive index. The imaginary part $V_I=\text{Imag}[\mathcal{V}]$ characterizes the dissipation, i.e. the potential is gain (loss) when $V_I>0$ ($V_I<0$). 
In the local regime ($\sigma\sim 0$), the above equation can be rewritten as
\begin{eqnarray}\label{eq:local NLSE}
i\frac{\partial u}{\partial \zeta}+\frac{1}{2}\frac{\partial^2 u}{\partial \xi^2}-\mathcal{V}u
+\bar{g}_0|u|^2u=0,
\end{eqnarray}
where $\bar{g}_0$ denotes the effective interaction strength. The local interaction is similar to the conventional Kerr nonlinearity, although the nonlinearity is much stronger.

By using the Madelung transformation, one can obtain hydrodynamic equations for the light intensity $\rho(\xi,\zeta)$ and flow velocity $v=\partial \phi/\partial \xi$, which can be further cast into the diagonal Riemann form after neglecting the quantum pressure, reading as    
\begin{subequations}\label{diagonal Riemann equation}
\begin{align}
&\frac{\partial r_1}{\partial \zeta}+c_1\frac{\partial r_1}{\partial \xi}=d_1,\\
&\frac{\partial r_2}{\partial \zeta}+c_2\frac{\partial r_2}{\partial \xi}=d_2.
\end{align}
\end{subequations}
Due to the existence of the dissipative potential, the right-hand side of the Riemann equation is no longer zero but is related to the dissipative potential, i.e. $d_{1,2}=\pm (r_1-r_2)V_I/2$.

\begin{figure}
\centering
\includegraphics[width=0.5\linewidth]{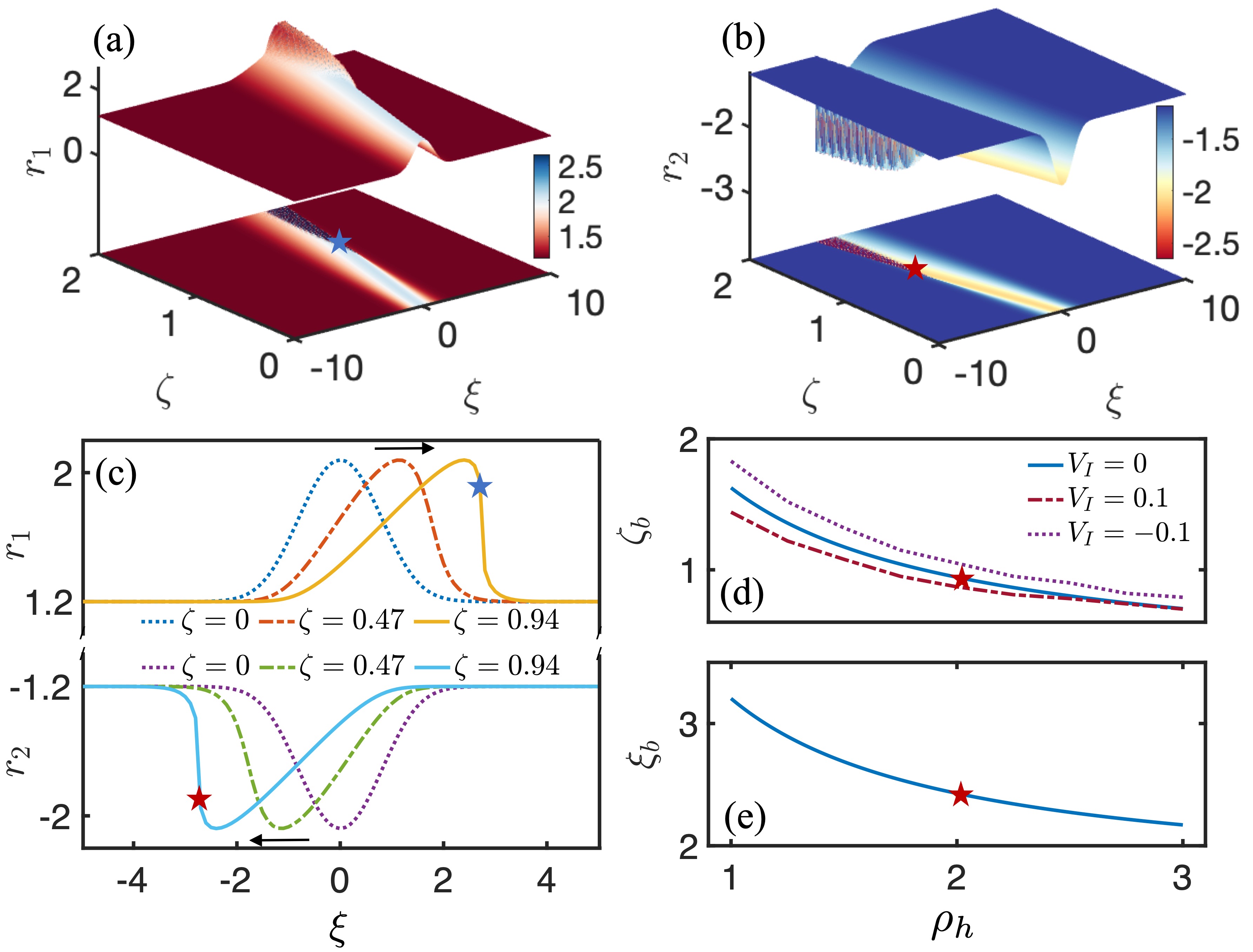}
\caption{\footnotesize (a) Right-moving and (b) left-moving Riemann waves. The stars represent the  breaking position of the wave, after which a shock wave is formed.
(c) Riemann waves $r_1$ and $r_2$ as function as $\xi$ for $\zeta=0,\,0.47,\,0.94$. The stars represent the  breaking position.  Upper panel: The right-moving part $r_1$. Lower panel: The left-moving part $r_2$. 
(d) Breaking point in the $\zeta$ direction ($\zeta_b$) as a function of the hump peak intensity $\rho_h$. The star marks the breaking point corresponding to panels (a) and (b).
(e)~The same as (d), but for the breaking point in the $\xi$ direction ($\xi_b$). Figures are adapted from Ref.~\cite{Qin2024PRA}.
}
\label{Ch3_Riemann_wave}
\end{figure}

Numerically solving the Riemann equation without dissipative potential with the initial condition $\rho(\xi,0)=\rho_b+\rho_h \exp\left[{-\xi^2/\xi_0^2}\right]$ and $~v(\xi,0)=0$. Figure~\ref{Ch3_Riemann_wave}(a) and (b) show the propagation of the right- and left-moving Riemann waves, respectively.
The wave is stable before the shock onsets. After a critical distance (marked by stars on the figure) the shock forms, i.e. breaking points. These points are symmetric for the left and right-moving solutions, as depicted in Fig.~\ref{Ch3_Riemann_wave}(c). 

Wave breaking corresponds to the occurrence of a gradient catastrophe, for which $|\partial r_{1,2}/\partial \xi|\rightarrow \infty$. 
Wave breaking occurs at a distance $\zeta$ such that~\cite{isoard2019PRA}
\begin{align}\label{wave-breaking distance}
\zeta_b \approx \frac{2\xi_0}{3(\rho^*-\rho_b)}\sqrt{\frac{\rho^*+\rho_b}{\bar{g}_0}}.
\end{align}
The breaking point as a function of $\rho_h$  is shown in the Fig.~\ref{Ch3_Riemann_wave}(d), which matches the numerical calculation well. The breaking point $\zeta_b$ is reduced when the hump intensity is increased. This relationship is useful for controlling the generation of shock waves. 
Moreover, the breaking point $\xi_b$ along the $\xi$ axis when the wave breaks along the $\zeta$ direction can be obtained~\cite{isoard2020EPL},
\begin{align}\label{xbreaking}
	\xi_b\approx c_s(\rho^*)\zeta_b+\xi_0\sqrt{{\rm ln}{\rho_h}-{\rm ln}[\rho^*-\rho_b]}.
\end{align}
Here, $c_s=\sqrt{\bar{g}_0\rho}$ is the local sound speed. The results are shown in Fig.~\ref{Ch3_Riemann_wave}(e), which agrees with the numerical calculation well.

\begin{figure*}[ht]
\centering
\includegraphics[width=0.83\linewidth]{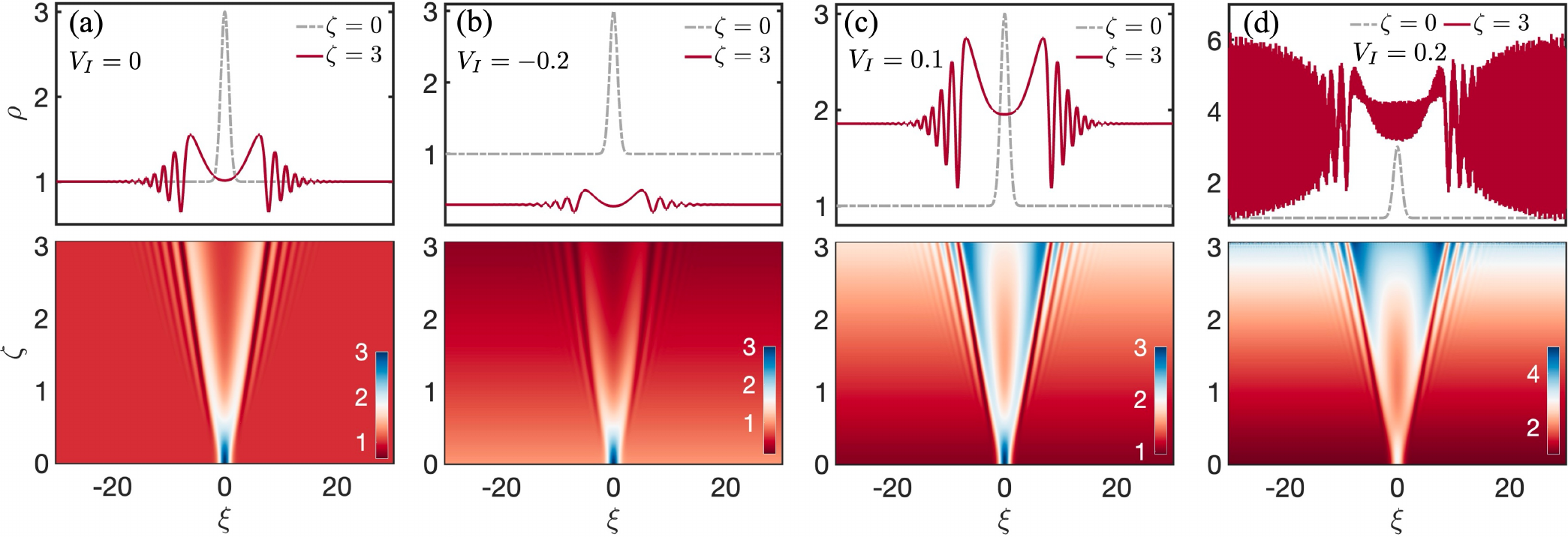}
\caption{\footnotesize
(a), (b), (c), and (d) show the propagation of shock waves with ${V_I}=0$ , ${V_I}=-0.2$ (loss), ${V_I}=0.1$, and ${V_I}=0.2$ (gain). Upper panel: The probe field intensity $\rho=|u|^2$ at $\zeta=0$ and $3$ (dotted-dashed gray and solid red lines). Lower panel: Wave propagation from $\zeta=0$ to $\zeta=3$. Figures are adapted from Ref.~\cite{Qin2024PRA}
}
\label{fig_propagation_local}
\end{figure*}

Including dissipation in the Riemann function, 
breaking points for $V_I=0.1$ and $-0.1$ are shown in Fig.~\ref{Ch3_Riemann_wave}(d).  It is found that the breaking point $\zeta_b$ decrease as the dissipative potential changes from loss to gain. These results highlight the importance of dissipative potential in the generation of shock waves. For example, the gain potential (i.e. $V_I>0$) accelerates the generation of shock waves, as the wave breaks earlier; see Figs.~\ref{fig_propagation_local}(c) and (d). The loss potential ($V_I<0$) slows down their generation; see Fig.~\ref{fig_propagation_local}(b).



\section{Conclusion and outlook}

Research on nonlinear optics based on Rydberg-EIT systems merges strong interactions from atomic physics with coherent control from quantum optics, achieving a leap from classical nonlinear wave phenomena to quantum few-photon nonlinearities at extremely low optical powers. From the diverse array of solitons, vortices, and spatiotemporal bullets supported by focusing nonlocal nonlinearity to pattern formation and shock wave dynamics in the defocusing regime, this platform exhibits unprecedented richness and tunability of nonlinear wave phenomena.

The theoretical predictions summarized in this review are grounded in a well-established experimental platform---ultracold Rydberg atomic gases in a ladder type and inverted-Y type EIT configuration. This setup has been successfully implemented in numerous laboratories~\cite{pritchard2010cooperative,Sinclair2019PRR}, and key nonlinear effects, such as giant cross-Kerr nonlinearities with third-order susceptibilities 
$\chi^{(3)}$ reaching $10^{-8}~\rm{m}^2/V^2$
 , have already been measured experimentally~\cite{Sinclair2019PRR}. These values are many orders of magnitude larger than those in conventional nonlinear media, providing strong support for the feasibility of observing the soliton, vortex, and shock wave phenomena discussed in this work. Moreover, the magnitude and degree of nonlocality of the Kerr nonlinearity can be conveniently tuned by adjusting the principal quantum number of the Rydberg state, the atomic density, and the detunings of the coupling lasers, offering experimentalists flexible control over the system parameters.

Despite this promising outlook, several technical and physical challenges remain to be overcome in translating these theoretical predictions into experimental realities. First, stability of high-D solitons. In (2+1)D and (3+1)D geometries (such as light bullets and vortex solitons), self-focusing media typically suffer from catastrophic collapse or strong instability. Although the nonlocal nonlinearity of Rydberg gases can, in principle, arrest collapse and stabilize these structures, achieving the precise balance between nonlinearity, diffraction, and dispersion remains experimentally demanding. Second, 
weak-light signal detection. While the system's ability to generate strong nonlinearities at ultra-low light intensities is a key advantage, it also poses detection challenges---producing and measuring nonlinear signals with sufficient signal-to-noise ratio while suppressing background noise from spontaneous emission and stray fields requires advanced techniques such as single-photon counting and homodyne detection. Third, 
Rydberg atoms are extremely sensitive to electric fields and collisions, so decoherence and dephasing must be carefully minimized through techniques such as high-vacuum environments and electric-field shielding. Overcoming these challenges will require careful parameter optimization and continued advances in cold-atom experimental techniques.

Beyond these classical nonlinear wave phenomena, recent work has extended the investigation to nonparaxial regimes. For instance, using the nonparaxial approximation, the spin-orbit interaction and light propagation properties can be studied based on the Rydberg long-range nonlocal interaction~\cite{Qin2026CPL}.

Several future research directions warrant attention. First, expansion to the quantum level: Although this review focuses on classical nonlinear optical phenomena, the ultimate potential of Rydberg-EIT systems lies in single-photon-level nonlinearities and deterministic photon-photon interactions~\cite{Pritchard2013in}. Single-photon transistors, single-photon switches, photonic logic gates, and deterministic sources of entanglement are ideal objectives in this field~\cite{Shi2022NC,Dudin2012,Hao2019SC,Gorniaczyk2014,Baur2014}. Second, hybrid system integration: coupling Rydberg atomic ensembles with platforms such as Bose-Einstein condensates, trapped ions, nanophotonic structures, or mechanical oscillators can leverage their respective strengths to realize novel hybrid quantum functionalities~\cite{Ashley2018}. Third, introduction of topological and gauge physics: In recent years, theoretical proposals have emerged for using Rydberg-EIT systems to simulate synthetic gauge fields and topological physics~\cite{Hu2026CPL,Wu2022PRR}; the confluence of nonlinearity and topology may give rise to new types of topological nonlinear modes~\cite{Khazali2022discretetimequantum}. Fourth, application-oriented functional devices: directions such as all-optical switches, beam controllers, and precision microwave electrometry are progressing from proof-of-principle demonstrations to performance optimization~\cite{Naseri2019}. As cold atom experimental techniques continue to advance and theoretical descriptions deepen, Rydberg atom nonlinear optics will remain at the forefront of strongly correlated photonic physics, paving new avenues for quantum information processing and manipulation of nonclassical light fields.


\section*{Acknowledgment}

L.Q., Z.S and C.H. acknowledge the National Natural Science Foundation of China (NSFC) under Grants Nos.~12404377, 12304357, 12374303. C.H  acknowledges National Key Research and Development Program of China (Grant No. 2022YFA1404202), and Shanghai Municipal Science and Technology Major Project (Grant No.~2019SHZDZX01).
Z.S  acknowledges Natural Science Foundation of Hubei Province (2023AFB352, 2026AFB581),
Outstanding Young and Middle-aged Scientific Innovation Team of Colleges and Universities of Hubei Province (T2023012), and Hubei University of Automotive Technology (BK202210). X.Z. acknowledges support from the Key Scientific Research Project of Colleges and Universities in Henan Province (26B140007) and the Key International Cooperation Project in Henan Province (Grant No. 261111521200). W.L. acknowledges support from the EPSRC through Grant No. EP/W015641/1. G.H acknowledges support from the Project Cultivation Fund of Fuyao University of Science and Technology
(Grant No. PF2025-A06).

\bibliographystyle{iopart-num.bst}
\bibliography{references}





\end{document}